\documentstyle[aps,epsf]{revtex}

\begin{document}
\draft

\title{ Renormalization Group and Perfect Operators for Stochastic
Differential Equations}

\author{ Qing Hou$^1$, Nigel Goldenfeld$^1$ and Alan McKane$^2$ }

\address{
$^1$Department of Physics,
University of Illinois at Urbana-Champaign, 1110 West Green Street,
Urbana, IL. 61801
}

\address{
$^2$Department of Theoretical Physics, University of Manchester,
Manchester M13 9PL, England
}

\maketitle

\begin{abstract}

We develop renormalization group (RG) methods for solving partial and
stochastic differential equations on coarse meshes.  RG transformations
are used to calculate the precise effect of small scale dynamics on the
dynamics at the mesh size.  The fixed point of these transformations
yields a perfect operator --- an exact representation of physical observables
on the mesh scale with minimal lattice artifacts. We apply the formalism to
simple nonlinear models of critical dynamics, and show how the method leads
to an improvement in the computational performance of Monte Carlo methods.

\end{abstract}

\pacs{PACS Numbers: 05.10.Cc, 05.10.-a, 05.10.Ln, 64.60.Ht}
\vspace{0.3in}

\section{Introduction}

The purpose of this paper is to introduce numerical methods that
avoid unnecessary discretization --- or, over-discretization ---
purely for the purpose of obtaining adequate accuracy.  An important
and classical example of this is large eddy simulation in the modeling
of turbulent flows.  Many large scale flows of engineering, geophysical
or atmospheric interest contain many length scales down to the
dissipation scale, yet it is large scale drag that one wants to
compute.  In such a situation, it is wasteful and undesirable to expend
computer time on details that are of no intrinsic interest.

The approach outlined in this paper builds upon our previous
work\cite{gmh} to use RG methods to integrate out the dynamics one
wishes to ignore, so that numerical methods can instead focus on the
appropriate scale of interest.  This is not trivial because of scale
interference: the nonlinear amplification of the effect of small scale
dynamics, which contaminates and eventually pollutes the large scale
dynamics. There are several distinct facets to this problem.

First is the representation of the small scale dynamics as a stochastic
field that acts on the coarse-grained degrees of freedom.  As discussed
in our earlier paper, this inevitably leads to non-locality.  We will
see here that it is possible not only to coarse-grain individual
operators, as in ref. \cite{gmh} but also to coarse-grain at the level
of the governing differential equation.  This leads to a theory that is
non-local in space and time.  This applies to systems with a finite
number of degrees of freedom, as well as spatially extended systems, which
are the main focus of our work here.

Second, the representation of the theory on the lattice can be improved
by systematically integrating out the small scales, leading to an
effective theory that has no (or few) residual discretization
artifacts.  This is referred to as a \lq\lq perfect theory" in the
literature.  We demonstrate how this arises and exhibit this feature by
calculating the dispersion relation of the effective theory in the
perfect representation.

Our work is related to that of Chorin and co-workers
\cite{ckk1,ckk2,ckl,chk,bcc} who use optimal prediction methods to
treat the lack of resolution of small scales.  The main differences are
that they assume that the small scales are initially in thermal equilibrium,
and also that they do not attempt to remove lattice artifacts. There has
also been an attempt to use similar methods in the study of isotropic
turbulence\cite{lang_mos}.

Our work grew out of attempts to improve lattice gauge theory,
pioneered by the paper of Hasenfratz and Niedermeyer.  For a review of
this body of work, the reader is referred to the review article by
Hasenfratz\cite{hasen}.  In addition to the work in ref. \cite{gmh},
there have been two attempts\cite{kw,hauswirth} to solve differential
equations using perfect operators. As we will see below, it is not
enough to perfectly coarse-grain the individual operators appearing in
a partial differential equation: once there is a non-infinitesimal time
step, coarse-graining introduces memory effects, so that the entire
differential equation must be represented as coarse-grained in
space-time.  In addition, it should always be remembered that there is
no unique perfect operator for a given differential operator.  A
specification must be made of the microscopic probability distribution
for the small-scale degrees of freedom.  These papers implicitly impose
a Gaussian free field theory distribution on the small-scale degrees of
freedom.  The methods given in the present article are more general,
and make no such assumption, explicit or implicit.

Let us now introduce the problem of removing lattice artifacts.
Suppose the dynamics of a spatially extended system is described by a
partial differential equation (PDE), which yields the solution $u(x,
t)$. A standard procedure is to sample $u(x, t)$ at points ${x_i,
t_j}$, which are equidistant with spacings $\Delta x$ and $\Delta t$,
and find a discretized form of the PDE that is devised to approximate
the values $u_{i,j} \equiv u(x_i, t_j)$. The requirement is that in the
continuum limit, the sequence $u_{i, j}$ converges to $u(x, t)$. The
conventional way of discretizing the PDE is to approximate
differentiations with finite differences.

The disadvantage of this {\it uniform sampling} (US) approach is that
one is forced to reproduce as faithfully as possible all the detail and
fine structure of the solution, even on a scale that may be of no
interest or worse, beyond the regime of applicability of the
differential equation itself. This has two consequences:

\begin{itemize}
\item a small grid size $\Delta x$ must be used, which implies many grid
points must be calculated and stored;
\item for dynamic problems,
a small time step $\Delta t$ is implied by the small $\Delta x$, either
for reasons of accuracy or stability of the numerical method.
\end{itemize}

As a result, there is a huge computational cost associated with this
conventional numerical scheme, which makes the study of problems such
as critical dynamics and pattern formation very difficult to carry out.
There is a need for improved, physically motivated methods for
numerical experiments.

The purpose of a numerical simulation is to study the macroscopic properties
of a physical system. Different microscopic dynamics may be related, via
{\it coarse graining} (CG), to the same macroscopic dynamics that
defines a universality class. Often CG means the local averaging of a
continuous variable,
\begin{equation}
U(X) = \int^{\Lambda/2}_{-\Lambda/2} dx\, u(X + x)\,,
\label{eqn:coarsegrainvar}
\end{equation}

\noindent
where $u(x)$  is the continuum variable, $U(X)$ is its coarse-grained
counterpart and $\Lambda$ is the coarse graining length scale. Instead of
focusing on the small scale degrees of freedom, we should determine and
use the coarse grained description of the system appropriate at the
macroscopic scale.

One of these physics-motivated numerical methods is the cell dynamical
scheme\cite{yoshicds}, in which a discrete description of the system
dynamics is obtained directly from considerations of the underlying
symmetry and conservation laws. It has been successfully used to tackle
problems such as asymptotic scaling behavior in spinodal
decomposition\cite{yoshi} and the approach to equilibrium in systems
with continuous symmetries, such as XY magnets\cite{mg3D} and liquid
crystals\cite{zap}. There have also been attempts at using the RG in
dynamic Monte Carlo simulations\cite{grant1,grant2}.

To investigate what is required to obtain a coarse-grained dynamic
description, suppose that we denote the coarse-graining operator at scale
$\Lambda$ by the symbol $C_\Lambda$, which transforms $u(x,t)$ to $U(X,t)$.
Then conceptually we need to find the operator $L_\Lambda$ which connects
$U(X,0)$ and $U(X,t)$ given the microscopic time evolution operator $L$
connecting $u(x,0)$ with $u(x,t)$, as shown schematically in the
commutativity diagram below:
\begin{eqnarray*}
u(x,0) &\quad\mathop{\longrightarrow}\limits^{L}\quad &u(x,t)\\
C_\Lambda\Bigg\downarrow &&C_\Lambda\Bigg\downarrow\\
U_\Lambda(X,0) &\quad \mathop{\longrightarrow}\limits^{L_\Lambda}\quad &
U_\Lambda(X,t)
\end{eqnarray*}

Notice that there is not a unique choice of $C_\Lambda$. The usual
choice is local averaging. In principle, other operators can be used,
such as the majority rule scheme used in the coarse graining of Ising
spins in thermal equilibrium. Once a coarse-graining operator
$C_\Lambda$ has been defined, there should be a unique prescription to
obtain $L_\Lambda$\cite{foot_1}. In this paper, coarse graining is
understood to mean local averaging. Later, stochastic coarse graining
will be introduced as a variant of the simple local averaging. In the
development of the theory of perfect operators a parameter, denoted
here by $\kappa_0$ (see section IIIB), naturally arises, which
characterizes the nature of the coarse-graining procedure. The form
(\ref{eqn:coarsegrainvar}) is only appropriate if $\kappa_0$ is
infinite; if $\kappa_0 < \infty$, additional noise terms are generated
which reflect the reduction in the number of degrees of freedom in the
system. As already stressed in our earlier paper\cite{gmh}, it is
inconsistent to work with a perfect operator with $\kappa_0 < \infty$
and to use the $\kappa_0 = \infty$ form (\ref{eqn:coarsegrainvar}) as
some authors\cite{kw,hauswirth} have done. We also see no reason why
coarse-grained equations should be derived by varying a coarse-grained
action in the absense of a small parameter, which is the starting point
of these authors. Instead we begin with a dynamics which is
intrinsically stochastic and study the effect of CG on this system. The
well-known path-integral formulation of such equations may then be used
to carry out the CG: there is no need to invoke a variational
principle.

We need to consider the appropriate coarse-graining scale.  Two
situations are possible here.  In the first, we suppose that the
solution we wish to obtain has a natural scale $\Lambda$ below which
there is no significant structure.  In that case, our goal is to avoid
having to over-discretize the problem merely in order to attain the
accuracy of the continuum limit.  Thus, we would like to be able to use
as large a value for the grid spacing $\Delta x$ as possible without
sacrificing accuracy.  In the second situation, there is no such obvious
scale, or at least, it is not known {\it a priori}, but the
computational demands are so large that it is simply not feasible to
work with a grid spacing $\Delta x$ smaller than some size $\Lambda$. In this
case, we would like to minimize in some sense the artifacts that must
inevitably arise.

The first situation is more straightforward because the only issue is speed
of convergence to the continuum limit: there is no explicit discarding of
important dynamical information. In the second situation, one is making an
uncontrolled and potentially severe truncation of the correct dynamics. One
has to ask: can one model the neglected unresolved scales as effective
renormalizations of the coefficients in the original PDE? Are the neglected
degrees of freedom usefully thought of as noise for the retained large-scale
degrees of freedom?  And how can any available statistical information on the
small-scale degrees of freedom be used to improve the numerical solution for
the large-scale degrees of freedom?

The plan of the paper is as follows. In section II we set up the
coarse-graining algebra, which forms the basis of our approach, using
the path-integral formulation of stochastic dynamics as our starting point.
This formalism is then used in section III to obtain the perfect operator
for dynamics governed by linear operators. Section IV describes the
results of numerical simulations using the perfect operator with
Langevin dynamics and section V using the Monte Carlo approach. A range of
issues is discussed, from applications of the method to the diffusion
equation and nonlinear model A dynamics to the question of the truncation
of perfect operators required when carrying out simulations. Our conclusions
are presented in section VI and the structure of the coarse-graining algebra
is discussed in an appendix.

\section{Coarse Graining in the Path Integral Formulation of Langevin
Dynamics}

In this section, we derive the path integral formulation of the
Langevin dynamics and present the general framework under which the perfect
linear operator is derived. The analysis is applicable to both PDEs and
stochastic differential equations. For simplicity, we study a system whose
dynamics is described by a stochastic differential equation (SDE) with the
following form,
\begin{equation}
\frac{\partial\phi(x, t)}{\partial t} = - f(x, t; \{\phi\}) + \eta(x, t)\,,
\label{eqn:c3_1PDE}
\end{equation}

\noindent
where $\phi$ is a field, $f$ is the forcing term (it can depend on $\phi$
and/or its spatial derivatives), and $\eta$ is a white noise.

It is convenient to regularize the problem on a (fine) $N \times N'$ lattice
with grid size $\Delta x$ and $\Delta t$ in the space and time directions,
respectively. In the lattice picture, all variables in the original PDE are
vectors of functions of discrete space $x = i \Delta x$ and time
$t = j \Delta t$ where $i \in [0, N-1]$, $j \in [0, N'-1]$. We define
$g(i \Delta x, j \Delta t) \equiv g(i, j)$ and denote the space-time volume
element $\Delta x \Delta t$ by $\Delta V$. The noise satisfies
$\langle\eta(i, j)\rangle = 0$ and $\langle \eta(i, j)\,\eta(i', j')
\rangle = \frac{\Omega}{\Delta V}\,\delta_{i, j}\,\delta_{i', j'}$ where
$\Omega$ is the noise strength and $\delta_{i, j}$ is the Kronecker symbol.
Given the system is in the state $\phi_0$ at time $t_0$, the probability
that the system will be in state $\phi_1$ at time $t_1$ is given by\cite{PI},
\begin{equation}
P(\phi_1, t_1 | \phi_0, t_0)  = \int  D\phi\, D\eta\, \exp\{-\frac{\Delta V}
{2 \Omega}\sum^{N, N'}_{i, j}[ \eta^{2}(i, j) -
\frac{\Omega}{\Delta x} \partial_{\phi} f] \} \,
\delta(\eta - \partial_t{\phi} -  f(\phi))\,,
\label{eqn:c3_1pi}
\end{equation}

\noindent
where the integration is over all configurations beginning at $\phi_0$ and
ending at $\phi_1$. We can use this path-integral formula to determine the
dynamics followed by the coarse grained or uniform sampled variable.

By a {\it discretization scheme}, we will mean a process made up of a series
of {\it magnifying operations} which lead from a microscopic description of
a system to a macroscopic description on a lattice. These magnification
operations are, by default, magnification of a length scale by a factor of
2. Coarse graining and uniform sampling are both special cases of a
discretization scheme.

Suppose a system is specified by the values of a function $f$, such as a
field configuration, on a fine lattice with 2N grid points
$x = (x_1, x_2, \cdots, x_{2 N})$ separated by grid size $\Delta x$. One
step (level) of coarse graining is defined as local averaging of the
function's values at every two neighboring sites.
\begin{equation}
\bar{f}_n = \frac{1}{2}(f_{2n-1} +  f_{2n}) \ \ \ , \ \ \
\tilde{f}_n = \frac{1}{2}(f_{2n} -  f_{2n-1}).
\label{eqn:c3_2CGf}
\end{equation}

\noindent
Vector $\bar{f}$ is the coarse grained version of $f$, while
$\tilde{f}$ stores the detailed information that is lost after coarse
graining. After one level of CG, the system is described by a new function
$\bar{f}$ on a coarser lattice with $N$ grid points separated by twice the
original grid size of $\Delta x^{M} = 2 \Delta x$, where the superscript $M$
indicates ``magnified value''. We define $2N \times N$ projection matrices
$\bar{\hat{R}}$, $\tilde{\hat{R}}$ such that

\begin{eqnarray}
\left\{\begin{array}{c}
f = \bar{\hat{R}}\,\bar{f} + \tilde{\hat{R}}\,\tilde{f} \\
\bar{f} = \bar{\hat{R}}^{-1}f \ \ , \ \ \ \tilde{f} =
\tilde{\hat{R}}^{-1} f\,.
\end{array}\right.
\end{eqnarray}

\noindent
These matrices act as projection and inverse projection operators between
the original functional space and the coarse grained functional space. They
facilitate an easier mathematical formulation. Many of the properties of
the matrices can be found in the appendix. If we are interested in an
operator $\hat{O}$ on the original grid, then it is possible to define
four corresponding operators on the coarse-grained grid, which we denote by
$\hat{O}_{A}, \hat{O}_{B}, \hat{O}_{C}$ and $\hat{O}_{D}$. For instance,
\begin{displaymath}
\hat{O}_D \equiv \bar{\hat{R}}^{-1}\hat{O}\bar{\hat{R}}\,.
\end{displaymath}

\noindent
The analogous definitions of $\hat{O}_{A}, \hat{O}_{B}$ and $\hat{O}_{C}$
are given in the appendix.

A similar algebraic scheme can be defined for the uniform sampling
transformation, where the projection operator samples every other point
and discards the rest:
\begin{equation}
{ \bar{f}}_n = { f}_{2n-1}, \mbox{ \hspace{0.3cm}}
{ \tilde{f}}_n = { f}_{2n}.
\end{equation}
\begin{equation}
{\bar{\hat{R}}}_{m, n} = \delta_{m, 2n-1}, \ \
{\bar{\hat{R}}}_{m, n} = \delta_{m, 2n}, \ \ \
m\in[1, 2N],\ \ n\in[1, N].
\end{equation}

\noindent
Using the notations listed above, we can write down the magnification
procedure in space for the $1+1$ dimensional version of (\ref{eqn:c3_1PDE}),
coarse graining in space only. The integrations over the $\phi$ and $\eta$
variables are decomposed into integrations over
$\bar{\phi}, \tilde{\phi}, \bar{\eta}$ and $\tilde{\eta}$ variables and
the $\tilde{\eta}$ integration carried out using the delta-function. The
remaining delta-function is replaced using the identity
$\delta(x) = a\,\delta(a \, x) = a\, \int dq e^{i a \, q \, x}/2\pi$. This
leads to a path integral, neglecting any constant factors, of the form
\begin{eqnarray}
P &=& \int { D\bar{\phi}} { D\bar{\eta}} { Dq}
\,\exp \{-\frac{\Delta V^{M}}{2 \Omega}\sum^{N, N'}[\frac{1}{c}\,\bar{\eta}^2
  - i { q}\, (\bar{\eta} - \partial_t{\bar{\phi}}) ]\} \times \nonumber\\
& &
\int D\tilde{\phi}
\, \exp\{-\frac{\Delta V^{M}}{2 \Omega}\sum^{N, N'}[\frac{1}{c}\,
(\partial_t{\tilde{\phi}} + { \tilde{f}})^2 + i { q} \,{ \bar{f}} -
\frac{\Omega}{\Delta x^{M}}( \partial_{\bar{\phi}} \bar{f} +
\partial_{\tilde{\phi}} \tilde{f})  ]\}\,,
\label{eqn:c3_2CGpi}
\end{eqnarray}

\noindent
where the constant $c$ is 1 or 2 for CG and US respectively due to their
different projection matrix properties and where
$\Delta V^{M}  = 2\Delta x \Delta t = 2 \Delta V$ is the magnified volume
element. The important point is that, in general, both ${ \bar{f}}$ and
${ \tilde{f}}$ are functions of ${\bar{\phi}}$ and ${\tilde{\phi}}$.

What we would like to do, is integrate over the ${ \tilde{\phi}}$ degrees
of freedom, carry out the $q$ integration and end up with a form similar
to the one we started with, but with new, renormalized, parameters. More
specifically, we would like the integration over ${ \tilde{\phi}}$ to give
a result of the form $\exp\{-\frac{\Delta V^{M}}{2 \Omega}(i q \, F -
\frac{\Omega}{\Delta x^{M}} \partial_{\bar{\phi}} F) \}$. Then we could
readily integrate over $q$ and compare the result with the path-integral form
to read off the evolution equation for the new coarse grained variable as
$\partial_t\bar{\phi} = - F(\bar{\phi}) + \bar{\eta}$. However, we would not
expect to be able to do this in general, and as usual in all applications
of the RG, an approximation scheme has to be developed alongside this
formalism in order to make any progress. There is, however, one case in
which the integrations can be carried out, and that is the linear case. We
therefore study this first, before returning to the nonlinear case later.

\section{Perfect Operator for Dynamics}

In this section, we will determine perfect operators of dynamics governed by
linear operators. We will find the fixed point flow of operators for the
diffusion equation under CG and US transformations. In addition, the
perfect operator in discrete space and time is obtained for the diffusion
equation and its properties discussed.

\subsection{Iterative Relations and Fixed Points in the linear case}

We begin by performing the magnifying transformation on the SDE
(\ref{eqn:c3_1PDE}) where $f$ is a linear function of $\phi$, that is,
\begin{equation}
\partial_t\,\phi = -{\hat{L}}\, \phi + \eta\,,
\label{eqn:c4_1force}
\end{equation}

\noindent
where $\eta$ is a white noise. Here ${\hat{L}}$ is a general linear operator
and contains spatial, but not temporal, derivatives. It is assumed to
possess inversion symmetry and translational invariance. For the diffusion
equation, ${\hat{L}}$  is the finite difference Laplacian operator with a
minus sign. The conventional choice is the central difference operator
${ \hat{L}}_{m, n} = (2\, \delta_{m, n} - \delta_{m, n+1} -
\delta_{m, n-1})/\Delta x^2$.

To obtain the dynamics of the coarse grained variable, we have to integrate
out the small length scale degrees of freedom in equation (\ref{eqn:c3_1pi}).
In the linear case, the Jacobian term is constant and so does not enter into
the analysis. Applying the projection matrices to equation
(\ref{eqn:c4_1force}), inserting the result into the path integral in
equation (\ref{eqn:c3_2CGpi}) and integrating out the $\tilde{\phi}$ and
$q$ degrees of freedom yields
\begin{equation}
P = \int  D\bar{\phi}  D\bar{\eta} \,\exp\{ - \frac{\Delta V^{M}}
{2 \Omega}\sum \left[ \bar{\eta}^{2} + (\bar{\eta} -
\eta^{M})\hat{Q}^{-1}(\bar{\eta} - \eta^{M})\right]\}\,,
\end{equation}
where $\eta^{M} \equiv \partial_{t}\bar{\phi} + (\hat{L}_{A} -
\hat{L}_{C}\hat{M}^{-1}\hat{L}_{D})\bar{\phi}$ and $Q \equiv \hat{L}_{C}
\hat{M}^{-1}(\hat{M}^{T})^{-1}\hat{L}^{T}_{C}$. Here the operator $\hat{M}$
is given by $\hat{I}\partial_{t} + \hat{L}_{B}$. Defining a new noise source
$\bar{\eta}' = \bar{\eta} - [\hat{I} + \hat{Q}]^{-1}\eta^{M}$ and carrying
out the integration over $\bar{\eta}'$ yields
\begin{equation}
P = \int  D\bar{\phi}  D\eta^{M} \,\exp\{ - \frac{\Delta V^{M}}
{2 \Omega}\sum \eta^{M} (\hat{I} + \hat{Q})^{-1} \eta^{M} \} \,
\delta(\eta^{M} - \partial_t \bar{\phi} - (\hat{L}_A - \hat{L}_C
\hat{M}^{-1}\hat{L}_D)\bar{\phi})\,.
\end{equation}

\noindent
Comparing with the form (\ref{eqn:c3_1pi}), it follows that the dynamic
equation satisfied by $\bar{\phi}$ is

$$\partial_t{\bar{\phi}} = - \hat{L}^{CG}\bar{\phi} + \eta^{M}\,,$$

\noindent
where $\hat{L}^{CG} \equiv  \hat{L}_A - \hat{L}_C \hat{M}^{-1}\hat{L}_D$.
The new noise source $\eta^{M}$ is no longer a white noise: it has a spatial
correlation as well as a time correlation:
\begin{equation}
\langle\eta^{M}\rangle = 0 \mbox{\ \ and }
\langle\eta^{M}(r, t)\eta^{M}(r', t')\rangle =
\frac{\Omega}{\Delta V^{M}}\,(\hat{I} + \hat{Q })(r - r', t - t')\,.
\end{equation}

Given that the noise source is no longer Markovian after the first step of
coarse graining, we need to start with a more general noise source in order
to iterate the coarse graining procedure. Define a general Gaussian noise
source with the following properties,
\begin{equation}
\langle\eta\rangle = 0 \mbox{ and } \langle\eta(r, t) \eta(r', t')\rangle =
\frac{\Omega}{\Delta V} \rho^{-1}(r - r', t - t')\,.
\end{equation}

\noindent
Repeating the above analysis, we find that the coarse grained dynamic
equation remains the same, however the coarse grained correlation matrix
is modified and is given by
\begin{equation}
(\hat{\rho^{CG}})^{-1} = \hat{L}_C \hat{M}^{-1}
\hat{\rho}_B^{-1}(\hat{M}^{-1})^T \hat{L}^T_C + \hat{\Gamma}\,(\hat{\rho}_A -
\hat{\rho}_C\hat{\rho}_B^{-1}\hat{\rho}_D)^{-1}\,\hat{\Gamma}^T\,,
\label{eqn:c4_1IterRho}
\end{equation}

\noindent
where $\hat{\Gamma} = \hat{I}+ \hat{L}_C \hat{M}^{-1}\hat{\rho}_B^{-1}
\hat{\rho}_D$. The presence of time derivatives in $\rho$ makes the noise
non-Markovian. In general, we should be careful about the boundary term in
this case\cite{mckanebray}. In particular, we need to specify corresponding
initial conditions for each time derivative generated through the iterative
relation.

The first term in $L^{CG} =\hat{L}_A - \hat{L}_C \hat{M}^{-1}\hat{L}_D $ is
{\it not} what we would naively choose as the Laplacian operator with a
coarse grained grid size $\Delta x^{M}$. Instead, the second term, which comes
from accounting for the influence of the integrated out small length scale
degrees of freedom, gives an important contribution to the coarse grained
operator and cannot be treated as a perturbation.

It is more convenient to examine the coarse graining in Fourier space (see
the appendix), where all matrices are now scalars dependent on wavenumbers
denoted by $k$ or $\kappa$, and frequencies denoted by $\omega$. We may
formally rewrite the iterative relation for $\hat{L}$ in Fourier space as,
\begin{equation}
\hat{L}^{CG}(\kappa) = \hat{L}_A(\frac{\kappa}{2}, \frac{\kappa}{2} \pm \pi)
+ \hat{L}_C(\frac{\kappa}{2}, \frac{\kappa}{2} \pm \pi)^2 /(i\omega +
\hat{L}_B(\frac{\kappa}{2}, \frac{\kappa}{2} \pm \pi))\,.
\end{equation}

\noindent
Each successive coarse graining procedure gives us a new operator which
weighs information from two different points of Fourier space, corresponding
to wave modes of different length scales, and puts them into a new point.
Even though the original linear operator contains only differentiation in
space, the new linear operator after one step of CG has a time
differentiation component as well. For $\omega = 0$, we can prove
analytically (and verify numerically) that the operator reaches a fixed point,
\begin{equation}
L(k) = \frac{4}{(\Delta x)^2} \sin^2\frac{k}{2} /( 1 - \frac{2}{3}
\sin^2\frac{k}{2})\,.
\end{equation}

\noindent
This is the perfect operator for $-\partial_x^2$ in one dimension. One might
hope that this operator can be recombined with $\partial_t$ and used in
the dynamic equation to give a perfect dynamics. It turns out that this is
in general {\it in}correct. The reason is that the iterative relation from
the path-integral calculation is a dynamic iterative relation with time
derivative in it. When one sets $\omega=0$, physically it translates into
the assumption that small scale degrees of freedom are enslaved by the
large scale dynamics. The small scale degrees of freedom instantaneously
adjust to the large scale ones which are kept after each magnifying
transformation. This is not physical.

Since we are only magnifying in space, the time differentiation is diagonal
in this phase space. We have the trivial relations,
$(\partial_t)_A = (\partial_t)_B = \partial_t$ and
$(\partial_t)_C = (\partial_t)_D = 0$. We define the full space-time
{\it evolution operator},
\begin{equation}
\hat{L}_{\omega} = \partial_t + \hat{L}  \mbox{\ \ such that \ \ }
\hat{L}_{\omega} \phi = \eta\,, \nonumber\\
\end{equation}
\noindent
and the {\it action operator},
\begin{equation}
H = \hat{L}^T_{\omega}\,\rho\,\hat{L}_{\omega} \mbox{\ \ such that \ \ }
\int  D\eta \,\exp\{ -\frac{\Delta V}{2 \Omega}\sum \eta\, \rho\,
\eta \}\, \delta(\eta - \hat{L}_{\omega}{\phi}) =
\exp\{ -\frac{\Delta V}{2 \Omega}\sum \phi\, H\, \phi \}\,,
\end{equation}
\noindent
and express the iterative relation in terms of $\hat{L}_{\omega}$ and $H$.
This leads to a simple form for the full iterative relation (see appendix),
\begin{eqnarray}
\left\{\begin{array}{rcl}
(\hat{L}_{\omega}^{-1})^{M} &=& (\hat{L}_{\omega}^{-1})_A
\label{eqn:TheIter1} \\
(H^{-1})^{M} &=& c \cdot (H^{-1})_A\,,
\label{eqn:TheIter2}
\end{array}\right.
\end{eqnarray}

\noindent
where the constant factor $c$ is 1 for CG and 2 for US. The second iterative
relation physically means that the coarse grained version of the two point
function of the true dynamics is preserved, if the coarse grained variable
is governed by the operator $\hat{L}_{\omega}$ with a non-Markovian noise
source $\rho$. The above iterative relations are readily generalized when
magnifications are carried out along the time direction.

\begin{figure}[ht]
\begin{center}
\leavevmode
\hbox{\epsfxsize=0.7\columnwidth \epsfbox{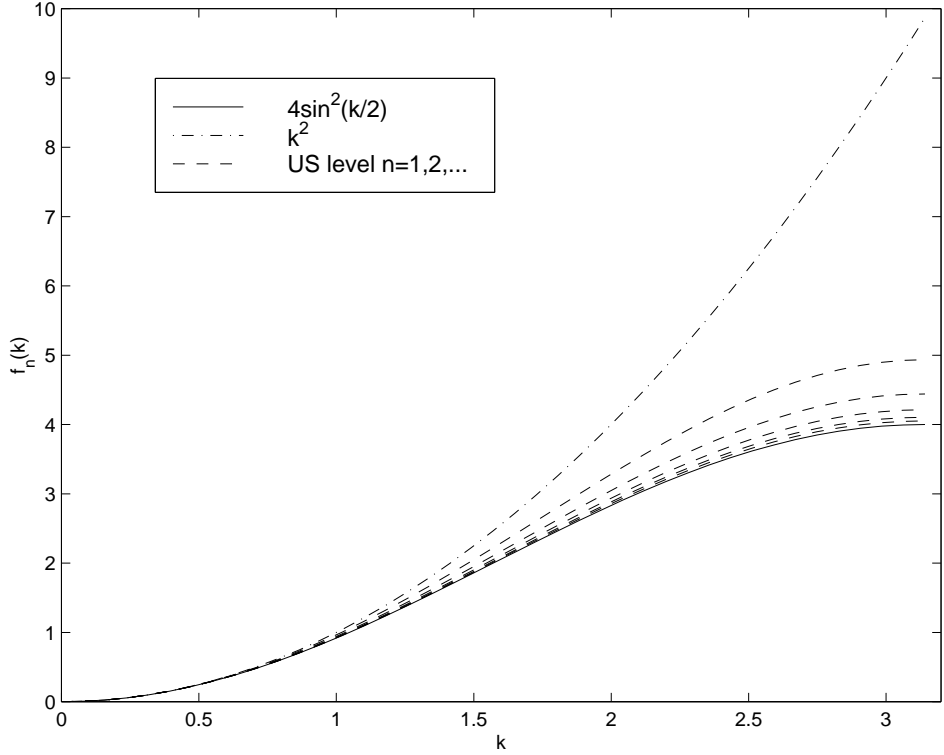}}
\end{center}
\caption{RG flow of the dynamics operator. In this case the starting point
is a microscopic Laplacian operator of the form
$L_{0, \omega}=i\omega + \frac{1}{\epsilon^2}k^2$.
The functional form of the $n^{\rm th}$ iterate of $L_{\omega}$ is,
$(L_{\omega, n})^{-1} = \frac{1}{2^n}\frac{\alpha_n}{i \omega \beta_n +
\frac{1}{\Delta x_n^2} f_n(k)}$.}
\label{fig:c4_2USfunc}
\end{figure}

We now wish to determine the fixed point solutions of the operators
$L_{\omega}$ and $H$ under their iterative relations. It can be shown that
the operators approach their fixed points exponentially fast as a function
of the number of iterative steps, irrespective of their detailed form at
the microscopic scale. The fixed point solutions are given below while the
exponential approach is illustrated in Figure \ref{fig:c4_2USfunc}.

We begin the the simpler case of US. Starting from a zeroth order operator
of the form
$L_{\omega, 0} = i\omega + \frac{4}{\epsilon}\sin^{2}\left(\frac{k}
{2}\right)$, appropriate for a description at the microscopic scale
$\epsilon$, after repeated US transformations we arrive at the operator
suitable for the length scale $\Delta x_{n} = 2^{n}\epsilon$. If the general
form of the US operator after $n$ iterations is written as
\begin{equation}
(L_{\omega, n})^{-1} = \frac{1}{2^n}\frac{\alpha_n}{i \omega \beta_n +
\frac{4}{\Delta x_n^2}\sin^2(\frac{k}{2})}\,,
\label{eqn:c4_2USLform}
\end{equation}
it is closed under iteration, given starting values $\alpha_{0}=\beta_{0}=1$.
The iteration relations are
\begin{equation}
\alpha_{n+1} = \alpha_{n}\left(1 + \beta_{n}\frac{i\omega\Delta x^{2}_{n}}
{2}\right)\ \ {\rm and}\ \
\beta_{n+1} = \beta_{n}\left(1 + \alpha_{n}\frac{i\omega\Delta x^{2}_{n}}
{4}\right)\,.
\end{equation}
These have a fixed point solution
\begin{eqnarray}\begin{array}{rclclcl}
\alpha_n &=& 1 + \frac{1}{6}(i\Theta_n) + \frac{1}{120}(i\Theta_n)^2 +
\cdots &=& \sum\,\frac{1}{(2i+1)!}(i\Theta_n)^i &=&
\frac{1}{\sqrt{i\Theta_n}}\sinh(\sqrt{i\Theta_n}) \nonumber \\
\beta_n &=& 1 + \frac{1}{12}(i\Theta_n) +  \frac{1}{360}(i\Theta_n)^2 +
\cdots &=& \sum\,\frac{2}{(2i+2)!}(i\Theta_n)^i &=&
\frac{2}{i\Theta_n}(\cosh(\sqrt{i\Theta_n})-1)\,.
\label{eqn:c4_2alphabeta}
\end{array}\end{eqnarray}
\noindent

where $\Theta_{n} \equiv \omega \Delta x^{2}_{n}$.

For the iterative process starting from, for instance, the microscopic
action operator $H_{0} = \omega^{2} + \left[\frac{4}{\epsilon^{2}}
\sin^{2}\left(\frac{k}{2}\right)\right]^{2}$, we take
\begin{equation}
H_n^{-1} = \frac{a_n + e_n \sin^2(\frac{k}{2})}{b_n^2\omega^2 +
(d_n\omega + \frac{4}{\Delta x_n^2}\sin^2(\frac{k}{2}))^2}\,,
\end{equation}
where $d_n$ and $b_n$ are the real and imaginary parts of $\beta$ in
$L_{n, \omega}$. The iteration relations for $a_n$ and $e_n$ are
\begin{eqnarray}
a_{n+1} & = & a_{n} + 2a_{n}d_{n}\frac{\Theta_n}{4} + (2a_{n} + e_{n})
(b^{2}_{n} + d^{2}_{n})\left(\frac{\Theta_{n}}{4}\right)^{2}\,, \nonumber \\
e_{n+1} & = & \frac{1}{4}\left(e_{n} + 4e_{n}d_{n}\frac{\Theta_n}{4}
- 2a_{n}\right)\,.
\end{eqnarray}

\noindent
The fixed point solutions are, setting $\theta_n = \sqrt{\Theta_n/2}$,
\begin{eqnarray}\begin{array}{rclclcl}
a_n &=& 1 + \frac{\Theta_n^2}{360}  + \cdots &=& \sum\frac{\Theta_n^{2i}}
{(4i+3)!}[4^{i+1}] -\frac{1}{2} e_n &=&
\frac{1}{4 \theta_n^3}(\sinh(2\theta_n) - \sin(2\theta_n))-\frac{1}{2} e_n\\
e_n &=& -\frac{2}{3} + \frac{4\,\Theta_n^2}{7!}  + \cdots &=&
\sum\frac{\Theta_n^{2i}}{(4i+3)!}[4 (-1)^{i+1}]&=& \frac{1}{\theta_n^3}
(\sinh(\theta_n)\cos(\theta_n)-\cosh(\theta_n)\sin(\theta_n))\\
b_n &=&1 - \frac{\Theta_n^2}{360} + \cdots &=& \sum\frac{\Theta_n^{2i}}
{(4i+2)!}[2 (-1)^i] &=& \frac{1}{\theta^2}\sinh(\theta_n)\sin(\theta_n)\\
d_n &=& -\frac{\Theta_n}{6}  + \cdots &=& \sum\frac{\Theta_n^{2i+1}}
{(4i+4)!}[2 (-1)^{i+1}] &=& \frac{1}{\theta^2}
(\cosh(\theta_n)\cos(\theta_n)-1)\,.
\end{array}\end{eqnarray}

\vspace{0.5cm}

We can now move on to the CG case. Here we parameterize the operators as
\begin{eqnarray}
L_{\omega, n}^{-1} &=& \gamma_n\,\frac{\Delta x_n^2}{4} +
\frac{\alpha_n}{i\omega\beta_n +
\frac{4}{\Delta x_n^2}\sin^2(\frac{k}{2})}\,, \nonumber \\
H_n^{-1} &=& f_n\,(\frac{\Delta x_n^2}{4})^2 + \frac{a_n + e_n
\sin^2(\frac{k}{2})}{b_n^2\omega^2 + (d_n\omega +
\frac{4}{\Delta x_n^2}\sin^2(\frac{k}{2}))^2}\,.
\label{eqn:c4_2CGlrhoForm}
\end{eqnarray}
\noindent

It is easy to see that CG shares the same $\beta$, $b$ and $d$ parameters
with US. The iteration relations for the other parameters are different.
However, one can obtain a relation between $\alpha^{CG}$ and $\alpha^{US}$,
namely,
\begin{equation}
\alpha^{CG}_{n} = \alpha^{US}_{n}\beta_{n} = \frac{1}{i\Theta_{n}}
\left(\alpha^{US}_{n+1} - \alpha^{US}_{n}\right)\,.
\end{equation}

\noindent
Using this relation we find
\begin{equation}
\gamma^{CG}_n = \frac{4}{i\Theta_n} (1 - \alpha_{n}^{US} )\,.
\end{equation}
\noindent

Therefore, the fixed point solution for $\alpha^{CG}$ and $\gamma^{CG}$
can be written in terms of that of $\alpha^{US}$, while the rest of the
parameters have fixed points
\begin{eqnarray}\begin{array}{llll}
a_n &= 1 - \frac{\Theta_n^2}{144} + \cdots &= \sum\frac{\Theta_n^i}
{(4i+5)!}[2^{2i+4}(RI - 1)] -\frac{1}{2} e_n &=  
(\cosh(\theta_n) - \cos(\theta_n))^2 Z_n\nonumber\\
e_n &= -1 + \frac{85\,\Theta_n^2}{3\cdot7!} + \cdots &=
\sum\frac{(-\Theta_n)^i}{(4i+5)!} [8 - 2^{4i+7}] &=
\frac{2}{\theta_n^4}(Z_{n+1} - Z_n)\nonumber\\
f_n &= \frac{2}{15} - \frac{16\,\Theta_n^2}{9!} + \cdots &=
\sum\frac{\Theta_n^i}{(4i+5)!}[16(-1)^i] &= -\frac{4}
{\theta_n^4}(Z_{n} -1)\,,
\end{array}\end{eqnarray}

\noindent
where $Z_n \equiv \frac{1}{2\theta_n}[\cosh(\theta_n)\sin(\theta_n) +
\sinh(\theta_n)\cos(\theta_n)]$, and $RI$ denote the average of real and
complex parts of $(\frac{3+i}{2})^{4n+5}$.

\subsection{Perfect Action Operator in Space-Time and Stochastic CG Scheme}

So far, we have only coarse grained the spatial degree of freedom and
obtained the corresponding perfect operators. In order to move on to
numerical calculations on a lattice, we also need to coarse grain the time
degree of freedom.

We focus on the perfect action operator
$H = \hat{L}^T_{\omega}\,\rho\,\hat{L}_{\omega}$ which is used later in the
space-time Monte Carlo calculations. Here we derive the fixed point
solution of $H$. We give a nearly closed form solution for $H(k, \omega)$ and
show that this operator gives a perfect dispersion relation as measured from
the time displaced two point function. The stochastic coarse graining scheme
is introduced, which modifies $H$ to give us an operator with reduced range
of interaction.

The iterative relation we developed previously does not hinge on whether CG
was carried out on the space or time axis. Therefore, we can use it to CG in
the time direction as well. Either one can start from a continuous
description and alternately CG in space and in time, or one can directly use
the perfect operator we developed previously and only CG from continuous
time. Now, there is another dimensionless parameter, namely the ratio of the
time scale over the characteristic time appropriate for a chosen length
scale. For the diffusion equation, it is $\Delta t/\Delta x^2$. We already
see the manifestation of this parameter in the perfect operator derived
earlier, where only the combination of the form $\omega \Delta x^2$ enters the
expressions. Therefore, there are two restrictions on how we apply the two
schemes. In the first case, we should CG twice in time direction for each CG
operation in the spatial direction, maintaining the value of the ratio
$\Delta t/\Delta x^2$ throughout the process. This means, for any reasonable
values of $\Delta t/\Delta x^2$ at the macroscopic side, we need to start
with a small $\Delta x$ and a very small $\Delta t$. In the second case, we
will not be able to maintain the ratio of $\Delta t/\Delta x^2$. Therefore,
the fixed point operator should be identified by iterating backwards. This
means, we repeat the iterative process many times starting from various
values of $\Delta t_n \equiv \Delta t/2^n$ and iterate $n$ steps. The fixed
point is identified as the operator which is (within tolerance) not changed
whether we start from $\Delta t_n$ or $\Delta t_{n+1}$. This method was
used in the previous section to calculate the fixed point operator form for
$H$ when the time frequency $\omega$ was nonzero. This reversed iteration
scheme is more powerful, since it can be generalized to other cases where
there are other dimensionless parameters, such as in the case of massive
fields.

The fixed point solution of a $d$-dimensional operator under the CG iterative
relation can be found using the techniques that have been described in this
paper. An alternative method, the so-called ``blocking from continuum'' can
also be used. In either case one finds\cite{bellwilson,wilsonblock,hn},
\begin{equation}
O_{\mathrm{FP}}({\bf k})^{-1} = \sum_{\bf l} O(({\bf k}+
2\pi{\bf l})/{\bf \Delta x})^{-1} \,\prod_{d=1}^{D}
\frac{4\sin^2(k_d/2)}{(k_d + 2\pi l_d)^2} + \frac{1}{\kappa_0}\,,
\label{eqn:c4_4sum}
\end{equation}

\noindent
where $O({\bf p})$ is the continuum spectrum of the operator and ${\bf l}$ is
a vector whose elements are of all possible integer values.

In the above equation, an extra constant term with a parameter $\kappa_0$ is
introduced. This term is important for obtaining a localized perfect operator
fit for numerical simulations\cite{hn}. To get this term, we modify the CG
procedure to be a {\it stochastic CG} operation, also called soft CG, instead
of hard CG, where an artificial noise term is introduced into the CG variable,
\begin{equation}
\phi^{S} = \bar{\phi} + \nu\,,
\end{equation}

\noindent
where $<\nu> = 0$ and $\langle\nu({\bf i})\,\nu({\bf i}')\rangle =
\frac{\Omega}{\kappa_0\Delta V}\delta_{{\bf i},{\bf i}'}$. Taking
$\kappa_0 \rightarrow \infty$, the hard CG case is recovered.

Now consider the diffusion equation for a massive field,
\begin{equation}
\partial_t \phi = \partial_x^2\phi - m\, \phi + \eta\,.
\end{equation}

\noindent
The continuum spectrum of $H$ is,
\begin{equation}
H = (\frac{\omega}{\Delta t})^2  + [(\frac{k}{\Delta x})^2 + m]^2
\mbox{\ \ \ where \ \ } \omega, k \in (-\pi, \pi)\,.
\end{equation}

\noindent
Defining the notation $x_l = x + 2\pi l$, we have,
\begin{equation}
\frac{1}{\Delta x^4}\,H^{-1} = \,\sum_{l, l'} \frac{1}{(k_l^2 + \mu)^2
  + r^2\,\omega_{l'}^2} \frac{4\sin^2(k/2)}{k_l^2}
\frac{4\sin^2(\omega/2)}{\omega_{l'}^2}  + \frac{1}{3\,r^2\,\kappa}\,,
\label{eqn:c4_4hamilFP}
\end{equation}

\noindent
where we defined parameters $\mu \equiv m \Delta x^2$ and
$r \equiv \Delta x^2/\Delta t$. To conform with notations used in quantum
field theories, we have defined $\kappa = \kappa_0\,\Delta t^2/3$.

The double summation is cumbersome to evaluate numerically due to its power
decaying behavior. By rewriting the factor
$\{ k^{2}_{l}[(k^{2}_{l} + \mu )^{2} + r^{2}\omega^{2}_{l'})] \}^{-1}$ as a
difference of two terms we can re-express the above formula as a sum of a
closed formed expression and an exponentially decaying expression. To do
so, it is convenient to introduce $\omega^{*}_{l} \equiv (k_l^2 + \mu)/r$ and
the function
\begin{displaymath}
G(k, \mu) \equiv \sum_{l}
\frac{4\sin^2(k/2)}{k_l^2 (k_l^2 + \mu)} =
\frac{1}{\mu}\{1 - \frac{(\sinh\sqrt{\mu})
( 1 - \cos k)}{\sqrt{\mu} ((\cosh\sqrt{\mu})-\cos k)}\}\}\,.
\end{displaymath}

\noindent
Then, after some simple algebraic manipulation, one finds
\begin{equation}
\frac{1}{\Delta x^4}\,H^{-1}  = -\partial_{\mu}G -
r\,\sin^2(\frac{\omega}{2})\, \partial^2_{\mu}G
+ 2r\,\sin^2(\frac{\omega}{2}) \sum_{l}
\frac{4\sin^2(k/2)}{k_l^2(k_l^2 + \mu)^3}
\frac{e^{-\omega^{*}_{l}} - \cos \omega}{\cosh \omega^{*}_{l} -
\cos \omega}  + \frac{1}{3\,r^2\,\kappa}\,.
\label{eqn:c4_4bettersum}
\end{equation}

\noindent
Now what remains of the summation is much easier to evaluate due to its
exponential decaying behavior.

From the above equation, we can obtain the dispersion relation implied by
such an operator. The two point function for a free field is
$S(k, \omega) = H^{-1}(k, \omega)$. Taking the discrete Fourier transform
of equation (\ref{eqn:c4_4bettersum}) back to real time gives the static
equal time structure factor,
\begin{equation}
S(k, t=0) = \sum_{l} \frac{4\sin^2(k/2)}{k_l^2}
\frac{1}{(\omega^{*}_l)^3}(\omega^{*}_l - 1 + e^{-\omega^{*}_l}) +
\frac{1}{3 r^2\,\kappa}\,,
\end{equation}

\noindent
and the time displaced two point function
\begin{eqnarray}
S(k, t \geq 1) &=& \sum_{l}\left\{ \frac{4\sin^2(k/2)}{k_l^2}
\frac{4\sinh^2(\omega^{*}_l/2)}{(\omega^{*}_l)^2} \right\}
\frac{1}{2\omega^{*}_l}e^{-\omega^{*}_l t} \nonumber\\
&=& \sum_{l} \frac{4\sin^2(k/2)}{k_l^2}\frac{1 - 2e^{-\omega^{*}_l} +
e^{-2 \omega^{*}_l}}{2(\omega^{*}_l)^3}e^{-\omega^{*}_l (t-1)}\,.
\label{eqn:c4_4decay}
\end{eqnarray}

\noindent
All dynamic modes are present, each with the correct decaying behavior and
with a prefactor (enclosed in curly bracket) due to coarse graining in space
as well as in the time direction. In principle, the decay rate should be
measured in the long time limit where all modes outside the first Brillouin
zone are negligible. However, for all practical purposes, the
$l \neq 0$ modes are negligible (or more precisely, the next significant mode
not degenerate with $l=0$) even for short times. For example, for
$k=\pi/2$, $\mu=0$, the amplitude of the next most significant mode
($l = -1$) is only $1.5\times 10^{-4}$ of that of the $l = 0$ mode.
Therefore, we can use the $t \geq 1$ values of the time displaced two point
function to evaluate the perfect dispersion relation for all the wave modes
with wavenumber within the first Brillouin zone.

 From $H^{-1}(k, \omega)$, we obtain the perfect operator coefficients
$H(r, t)$ in real space and time. Notice that ``the fixed point
of an operator'' actually means the fixed point of the dimensionless
operator. Consequently, operator coefficients for the perfect action operator
are actually those of $H\,\Delta x^4$. For practical reasons, we need to
adjust the parameter $\kappa$ for optimal locality. In one dimension,
$\kappa \approx 2$ and 6 are the best values for $\partial_x^2$ and
$\partial_x^4$ respectively. Therefore, we need to find a compromise. The
best scheme is to choose $\kappa=2$ such that the most significant couplings
lie within a rectangle area elongated along the $x$ direction. This way, the
total number of significant couplings is minimized.

\begin{figure}[bht]
\begin{center}
\leavevmode
\hbox{\epsfxsize=0.8\columnwidth \epsfbox{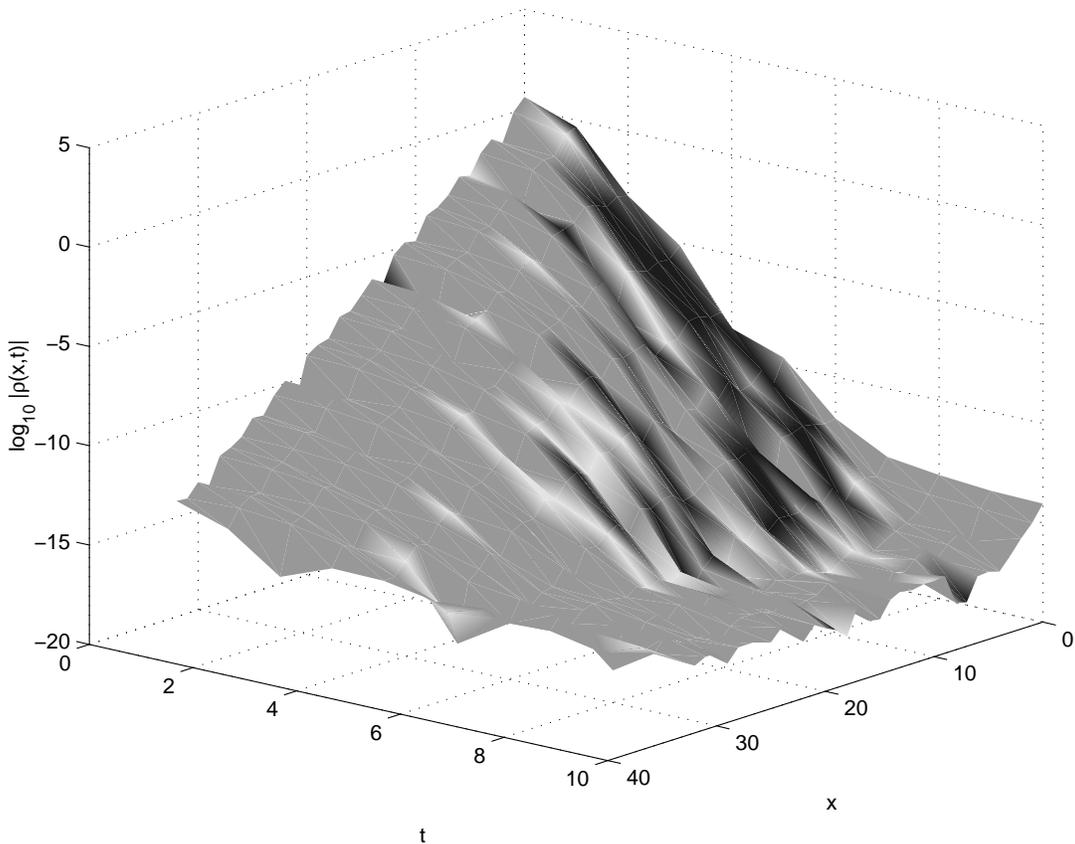}}
\end{center}
\caption{Surface plot of amplitude of perfect action operator coefficients
for the diffusion equation. The coefficients exponentially decay away from
the origin. The decay speed is slow along the $x$ direction. $\kappa = 2$,
$m=0$ and $\Delta t = \Delta x^2$.}
\label{fig:c4_4rho}
\end{figure}

The leading order coefficients of $H$ for $\kappa = 2$ and zero mass are
tabulated in Table \ref{tbl:c4_4rho} and are shown in
Figure \ref{fig:c4_4rho}.

\begin{table}
\caption{Sample coefficients of the perfect action operator for the diffusion
equation. $\kappa = 2$, $\mu=0$ and $\Delta x^2/\Delta t = 1$.}
\begin{center}
\begin{tabular}{|c|c|c|c|c|c|c|c|}\hline
$(t, x)$ & $H$ & $(t, x)$ & $H$ & $(t, x)$ & $H$ & $(t, x)$ & $H$\\ \hline
(0, 0) & 3.90458 & (0, 1) & -1.02978 & (0, 2) & -0.0421266 & (0, 3)
& 0.098042
\\ \hline
(0, 4) & 0.0291451 & (0, 5) & -0.00317113 & (0, 6) & -0.00407848 & (0, 7)
& $-7.35334 \times 10^{-4}$\\ \hline
(1, 0) & -0.464966 & (1, 1) & -0.278677 & (1, 2) & -0.0339122 & (1, 3)
& 0.0328692 \\ \hline
(1, 4) & 0.0148371 & (1, 5) & $-2.8286 \times 10^{-4}$ & (1, 6)
& -0.00204057 & (1, 7) & $-5.22915 \times 10^{-4}$ \\ \hline
(2, 0) & $-5.99324 \times 10^{-4}$ & (2, 1) & $-8.25418 \times 10^{-4}$ &
(2, 2) & $-3.68371 \times 10^{-4}$ & (2, 3) & $6.91673 \times 10^{-4}$
\\ \hline
(2, 4) & $8.04927 \times 10^{-4}$ & (2, 5) & $2.04135 \times 10^{-4}$ &
(2, 6) & $-1.27393 \times 10^{-4}$ & (2, 7) & $-9.25806 \times 10^{-5}$
\\ \hline
\end{tabular}
\end{center}
\label{tbl:c4_4rho}
\end{table}

\section{Numerical Simulation Using the Perfect Operator}

In this section, we discuss the application of the perfect linear operator
in numerical simulations of Langevin dynamics. We show that the perfect
operator should be decomposed into an {\it Up operator} and a {\it Down
operator} in order to obtain a correct equation with a finite number of high
order time derivatives. Without this decomposition, the truncation of the
perfect operator is highly non-trivial, if not impossible. For Langevin
dynamics, the dynamics of the non-Markovian noise is difficult to obtain
because it requires taking the square root of the noise correlation function.
Various numerical simulations were carried out using the truncated perfect
operator and other approximations, to illustrate the advantage of using coarse
grained variables as opposed to uniformly sampled variables in numerical
simulations. These, together with the limitations of this approach, are also
discussed.

\subsection{Perfect Operators in Langevin Dynamics}

Here we derive the perfect operators $\hat{U}$ and $\varrho$ appropriate for
Langevin dynamics. In the previous section, we obtained the iterative
relation for both $\hat{L_{\omega}}$ and $H$. From the latter, the
correlation function for the non-Markovian noise is obtained. The
discretized system follows the dynamics described by the PDE,
\begin{equation}
\hat{L_{\omega}} \phi = \eta\,,
\label{eqn:c5_1lang1}
\end{equation}

\noindent
where $\eta$ satisfies $\langle\eta(i, j)\,\eta(i', j')\rangle=
\frac{\Omega}{\Delta V}\rho^{-1}(i-i', j-j')$. This formally simple equation
is different from the usual Langevin dynamics in two respects: the
non-Markovian nature of noise and the presence of (in principle) infinite
orders of time derivatives in both $\hat{L_{\omega}}$ and $\rho^{-1}$.

The non-Markovian nature of the noise means that there is dynamics in the
noise variable. This is not surprising. In the path-integral calculation,
each CG step results in formally discarding small scale degrees of freedom.
But in fact, the small scale degrees of freedom are not entirely discarded.
Since the small scale dynamics is affected by the noise source as well as
the system dynamics at the coarse grained level, when the small scale degrees
of freedom are integrated out at each CG step, part of the small scale
dynamics is preserved by modifying the dynamics at the larger length scale
and by injecting dynamics into the noise. This is basically a feedback
effect.

Due to the non-Markovian nature of the noise, we need to write down the
dynamics followed by the noise,
\begin{equation}
\rho^{1/2} \eta = \eta_0\,,
\end{equation}

\noindent
where $\eta_0$ is a white noise satisfying $\langle\eta_0(i, j)\,
\eta_0(i', j')\rangle=\frac{\Omega}{\Delta V}\delta_{i, i'}\,
\delta_{j, j'}$. The matrix $\rho^{1/2}$ is the square root of $\rho$ in the
sense that the product of $\rho^{1/2}$ and its Hermitian conjugate gives
$\rho$. For instance, in Fourier space,
$\sqrt{\omega^2 + k^4} = i \omega + k^2$. There are in principle infinite
orders of time derivatives in $\rho^{1/2}$, just as in $\hat{L_{\omega}}$.

Naively, $\hat{L_{\omega}}$ can be obtained as a series expansion in
$\partial_t$ which is then truncated to certain order. This turns out not to
be the correct approach. Rather, we need to decompose the operator
$\hat{L_{\omega}}$ in the form of a numerator ($\hat{U}$) over a denominator
($\hat{D}$),
\begin{equation}
\hat{L_{\omega}} \equiv \hat{D}^{-1}\hat{U}\,,
\end{equation}

\noindent
where we write the denominator as an inverse operator. The distinction
between the numerator and denominator is easily seen in the fixed point
operator. We can eliminate the inverse operator by applying $\hat{D}$ on
both sides of equation (\ref{eqn:c5_1lang1}). Redefining the noise as
$\xi = \hat{D}\eta$ and denoting its correlation function
$\varrho^{-1}$, we have,
\begin{equation}
\hat{U}^{-1}\varrho^{-1}\hat{U}^T\,^{-1} = \hat{L}^{-1}\rho^{-1}
\hat{L}^T\,^{-1}\,.
\end{equation}

\noindent
The operators $\hat{U}$ and $\varrho$ are therefore equivalent to the older
pair of $\hat{L}$ and $\rho$ in the evolution of the discretized system.
Equation (\ref{eqn:c5_1lang1}) may be rewritten as
\begin{eqnarray}
\hat{U}_{\omega} \phi =\xi \mbox{\ \ \ \ with \ \ \ \ }\varrho^{1/2}
\xi = \eta_0\,.
\label{eqn:c5_1UDeqn}
\end{eqnarray}

\noindent
Using (\ref{eqn:c4_2CGlrhoForm}), and in the notation of the last section,
the perfect operators for the diffusion equation under the CG scheme
are
\begin{eqnarray}
\hat{U}_{\omega} &=&  i\omega\beta + \frac{4}{\Delta x^2}\sin^2\frac{k}{2}\,,
\nonumber\\
\varrho^{1/2} &=& \{\alpha + \gamma\,\sin^2\frac{k}{2} +
\gamma\,\beta\frac{i\Theta}{4}\}/\{a + e\,\sin^2\frac{k}{2} +
f\,b^2\,(\frac{\Theta}{4})^2 + f\,(d\,\frac{\Theta}{4} +
\sin^2\frac{k}{2})^2\}^{\frac{1}{2}}\,.
\label{eqn:c5_1UDCG}
\end{eqnarray}

\noindent
For analytical tractability, we used the closed-form solutions of the
operators available for discrete space but continuous time.

Unlike $\hat{L}$, the new evolution operator $\hat{U}$ can be expressed in a
clear and simple series expansion. The spatial part is simply the central
difference operator and the time part is a sum of all orders of time
derivatives with constant and fast decaying coefficients (see equation
(\ref{eqn:c4_2alphabeta})).

The operator $\varrho^{1/2}$ has a very complicated form. It has many high
order space and time derivatives, which in general are coupled. Series
expansion and truncation are necessary. To the first order in
$\Delta x^2$, we have for CG,
\begin{equation}
\varrho^{1/2} \approx 1 - \frac{1}{6}\,\sin^2(k/2) + c\,i\omega\,\Delta x^2 =
1 + \lambda^2\,\partial_x^2 - \tau \partial_t\,,
\end{equation}

\noindent
where $c = 1/6 - 1/\sqrt{720} = 0.129$, $\lambda \equiv \Delta x/\sqrt{24}$
and $\tau\equiv c\,\Delta x^2$. Therefore, the noise source is largely a
white noise. It has a correlation length of the order $\lambda$ and a
relaxation time of the order $\tau$. When the form of the operator is
obtained and truncated to a specified order, one can evolve the system
according to equation (\ref{eqn:c5_1UDeqn}).

Often periodic boundaries are used in the spatial dimensions. Therefore high
order spatial derivatives do not pose a problem on a lattice. Higher order
time derivatives, however, require a corresponding number of initial
conditions. This might pose a problem, especially for the non-Markovian
noise. If one is interested in equilibrium properties of the system, the
initial transient stage is not important. An initial condition with all
derivatives zero is fine. When one wants to study the initial transient
stage corresponding to a certain microscopic initial condition, one can
evolve the system using a fine mesh for $n$ steps under a conventional
numerical scheme, where $n$ is the highest order time derivatives. For each
step, one can coarse grain the microscopic configuration to the desired CG
level, insert the CG version of $\phi$ into equation (\ref{eqn:c5_1UDeqn}),
and the noise in the transient stage is obtained. This way, initial time
derivatives for both coarse grained $\phi$ and $\xi$ can be computed.

The calculation of the space-time discretized $\varrho^{1/2}$ can be quite
involved\cite{qingsqrt}. Since our main interest is in calculating
equilibrium properties of dynamic systems, we can take an alternative route,
namely Monte Carlo simulation, as discussed later. In this case, the
perfect action operator $H$ is all we need.

\subsection{An Example of Using Perfect Operator in Langevin Dynamics}

In this section, we present an application of the operator $\hat{U}$ to
the deterministic dynamics of the coarse grained variable governed by the
diffusion equation. The (truncated) perfect operator $\hat{U}$ gives superior
results for the evolution of the configuration. The relative advantage of
using the CG variable vs the US variable is also touched upon and will be
studied more closely in the ensuing section.

For simplicity, we truncate the series expansion of $\beta$ to the first
order to obtain an operator $\hat{U}$ with a second order time derivative.
Direct truncation is not appropriate when setting higher order terms to zero,
since we should adjust the remaining coefficients. Instead, we use the
operator at the first level of CG, starting from a central difference
operator. The coefficients for time derivatives higher than the second order
are identically zero. We have,
\begin{equation}
\frac{\Delta x^2}{16}\partial^2_t \phi_i + \partial_t \phi_i +
\frac{1}{\Delta x^2}(2\, \phi_i - \phi_{i+1} - \phi_{i-1}) = 0.
\label{eqn:c5_1UDeqnEg}
\end{equation}

\noindent
Suppose the system is periodic with length $L$. The initial condition has
modes down to length scale $\epsilon = L/M$ with $M$ being an integer, namely,
\begin{equation}
\phi(m, t=0) = \sum^{M/2}_{k=-M/2} e^{i 2\pi k\,m\epsilon/L}\phi_k\,,
\end{equation}

\noindent
where $\phi_k$ is the amplitude of the $k$th wave mode. We know analytically
the exact solution: by coarse graining the exact solution to a length scale
$\Delta x = L/N = p \,\epsilon$, we have,
\begin{equation}
\bar{\phi}(n, t) =
\sum^{N/2}_{q=-N/2} e^{i \frac{2\pi q\,n\Delta x}{L}} \sum^{p/2}_{i=-p/2}
\phi_{q + i\,N} \frac{\sin(\pi q\Delta x/L)}{p
\sin(\pi (q + i\,N)\epsilon/L)} e^{-(\frac{2\pi q\Delta x}{L})^2t}
\equiv \sum^{N/2}_{q=-N/2}
e^{i \frac{2\pi q\,n\Delta x}{L}} \bar{\phi}_{q,0}(t)\,,
\end{equation}

\noindent
where $\bar{\phi}_{q,0}(t)$ is the exact wave mode for the CG variable. This
equation gives us both $\bar{\phi}(n, t=0)$ and
$\partial_t\bar{\phi}(n, t=0)$. Now let us ask: what result would equation
(\ref{eqn:c5_1UDeqnEg}) yield on a lattice with grid size $\Delta x$, given
the CG initial conditions? We have $\bar{\phi}_q(t) = \sum^{p/2}_{i=-p/2}
C_{q,i}(t)\, \phi_{q + i\,N}$, where
\begin{equation}
C_{q,i}(t) = \frac{\sin(\pi q\Delta x/L)}{p \sin(\pi (q + i\,N)\epsilon/L)}
e^{-\omega_{-} t}\, \{ 1 +
\frac{ 1- e^{-\Delta \omega\,t}}{\Delta \omega}[\omega_{-} -
\frac{2\pi(q + i\, N)}{L^2}]\},
\label{eqn:c5_1cqi}
\end{equation}

\begin{figure}[htb]
\begin{center}
\leavevmode
\hbox{\epsfxsize=0.7\columnwidth \epsfbox{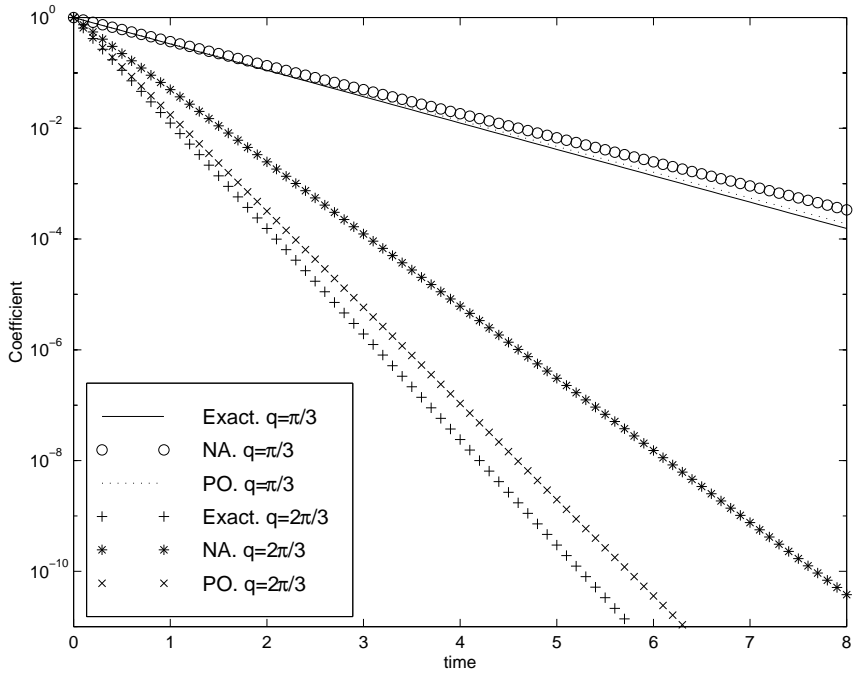}}
\end{center}
\caption{Decay of wave modes in the first Brillouin zone using PO and NA
equations vs the exact result. The decay rates for the PO scheme are closer to
the exact ones than the NA results. The coefficient is $C_{q,i=0}(t)$ without
the CG prefactor. Sample wavenumbers are $q=\pi/3$ and $2\pi/3$. }
\label{fig:c5_1coef0}
\end{figure}

\noindent
$\omega_{-} = \frac{16}{\Delta x^2}\sin^2(\frac{\pi q}{2N})$, and
$\Delta \omega = \frac{16}{\Delta x^2}\cos(\frac{\pi q}{N})$. For comparison,
the corresponding result from conventional numerical analysis
(NA), which is the same as just keeping the first-order time derivative in
$\hat{U}$, is
\begin{equation}
C^{NA}_{q,i}(t) = \exp\{-\frac{4 \sin^2(\pi q/N)}{\Delta x^2}t\}\,,
\end{equation}

\noindent
where the time evolution does not depend on $i$. The solution for modes
within the first Brillouin zone, i.e. $i=0$, is greatly improved as shown in
Figure \ref{fig:c5_1coef0}, where we have plotted the time evolution of the
coefficient $C_{q,i=0}(t)$ (without the prefactor due to CG) for selected
$q$ values. For small $q$, the $\Delta \omega$ dependent part in equation
(\ref{eqn:c5_1cqi}) is not important. A Taylor expansion clearly shows that
$\omega_{-}$ is closer to the true decay rate than the NA result.
\begin{figure}[htb]
\begin{center}
\leavevmode
\hbox{\epsfxsize=0.7\columnwidth \epsfbox{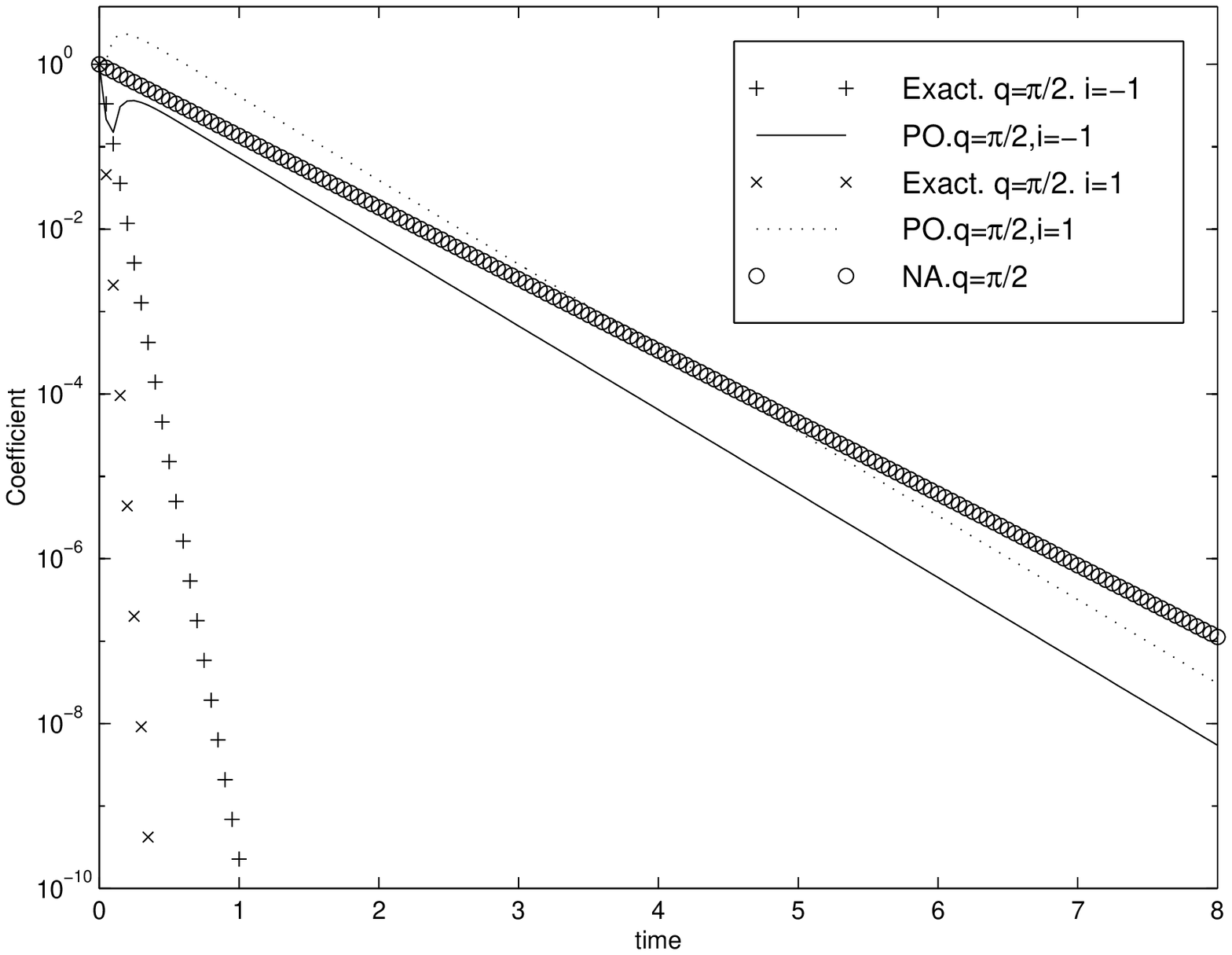}}
\end{center}
\caption{Decay of wave modes in the second $(i=-1)$ and third $(i=1)$
Brillouin zones using PO and NA equations vs the exact result. For the PO and
NA schemes, wave modes do not decay as fast as for the exact result. PO is
better for modes in the second Brillouin zone than NA and is also
advantageous for late times for modes in other Brillouin zones. PO results
are very close to the exact one at short times as indicated by the dip in the
plotted curve. The coefficient is $C_{q,i=0}(t)$ without the CG prefactor,
which reduces the importance of modes outside the first Brillouin zone.
Sample wavenumber is $q=\pi/2$. Notice that the NA result does not depend on
$i$.}
\label{fig:c5_1coef1}
\end{figure}

For very short times, the dynamics of all modes are correctly prescribed,
even for $i \neq 0$. This manifests itself as a dip in the short time region
of Figure \ref{fig:c5_1coef1} (subject to resolution in the time axis, the
dip for the $i=1$ mode is not discernible in the figure). For finite time,
modes outside the first Brillouin zone decay quickly in the
true dynamics (Figure \ref{fig:c5_1coef1}). In the PO result, the decay rate
is dependent on $q$, therefore these modes do not decay as fast as they
should do. But since the PO result also contains information on $i$, for
modes in the second Brillouin zone, the resulting dynamics are still closer
to the true one than the NA result. This is because we used the PO operator
$\hat{U}$ of one level CG. For higher wavenumbers, due to the $i$ dependent
term, there is an anomalous (negative) amplification of wave modes at the
initial transient stage which disappears later. Therefore, the power spectrum
of the configuration should die off quickly for modes with length scale
much less than $\Delta x$. In other words, we should not over-coarse-grain.
It follows that as we keep more and more terms in the perturbative series for
$\hat{U}_{\omega}$, the PO result will be close to the exact one for higher
and higher wave modes.

The prefactor
\begin{displaymath}
\frac{\sin(\pi q\Delta x/L)}{\sin(\pi (q + i\,N)\epsilon/L)}
\end{displaymath}

\noindent
modifies the contribution of each wave mode to the solution. This comes from
using the CG variable in the PDE and is very important in reducing errors
that arise from using the discretized PDE. For instance, although modes with
$q \approx 0$ decay very slowly in the PO result, their prefactor is close
to zero for $i \neq 0$, while they very quickly decay to zero in the true
dynamics.

Notice that US and CG share the same $\hat{U}$. In the US scheme, there is no
prefactor. Modes with $q \approx 0$ and $i \neq 0$ do not decay. If we use
the same equation as above, the prefactor for $ t > \Delta x^2$ is,
\begin{equation}
C_{q \approx 0,i}(t) \approx   1 - (\frac{i\pi}{2})^2\,.
\end{equation}

\noindent
For large $i$, it overstates the contribution of the mode to the solution
and is worse than NA. This imposes a stricter constraint than for the CG
scheme on the power spectrum of the configuration, and is the reason why CG
is a better scheme. This has been tested numerically on several model
dynamics\cite{gmh}.

\section{Space-Time Monte Carlo Simulation}

The path-integral formulation easily leads to a space-time Monte Carlo
simulation. We discuss issues related to truncating the perfect operator
such that it has a finite range of interaction. Numerical simulations are
carried out on the linear diffusion equation to test computational
efficiency of using the perfect operator, and on Model A dynamics to test
the merit of direct application of the perfect linear operator to nonlinear
dynamics.

In quantum field theories, many problems are formulated in terms of path
integrals. Numerical simulations usually employ the Monte Carlo method,
where due to space-time symmetry, time is simply treated as one of the
dimensions in a $d+1$-dimensional lattice. In statistical physics, when
dynamics is involved, evolving a Langevin equation is the norm. A typical
form of the equation contains a first-order time derivative, a diffusion term
and some nonlinear interaction. Time and space are not symmetric. However,
numerical simulation of a Langevin equation is not the only choice for
studying dynamics. We can also perform Monte Carlo simulation on a
space-time lattice\cite{zimmer}, similar to the approach adopted in quantum
field theories. The basis for such a calculation is the path-integral
formulation.
Starting from equation (\ref{eqn:c3_1pi}), and performing a trivial
integration over the noise to eliminate the delta function, we have
\begin{equation}
P = \int  D\phi \,\exp\{-\frac{\Delta V}{2 \Omega}\sum_{i, j}
[ (\partial_t{\phi} +  f(\phi))^2 - \frac{\Omega}{\Delta x}
\partial_{\phi} f]\}\,.
\end{equation}

\noindent
The cross term linear in $\partial_t$ results in a boundary term and does
not influence directly the calculation of $P$. Ignoring the Jacobian
contribution, we are left with a positive definite functional. We call the
term in the exponent the `action' for obvious reasons. For linear operators,
we know how to coarse grain the above expression. Integrating out the noise
in the above equation, we have
\begin{equation}
P = \int D\bar{\phi} \,\exp\{-\frac{\Delta V}{2 \Omega}\sum \phi \,H
\,\phi\}\,,
\label{eqn:c5_2CGpi}
\end{equation}

\noindent
where $H$ is the fixed point operator of the action operator. Working with
this path-integral formulation, we do not have to worry about taking the
square root of the noise correlation matrix as we would with the Langevin
equation.

In the following, we will look at a specific example of the linear theory,
namely the dynamics of a system described by the diffusion equation,
\begin{equation}
\partial_t \,\phi = \partial_x^2 \,\phi - m \,\phi + \eta\,,
\label{eqn:c5_2difEqn}
\end{equation}

\noindent
where $m$ is a constant which we will call the mass and $\eta$ is white noise
with strength $\Omega$. We have chosen a unit diffusion constant. In the
space-time Monte Carlo probability we use the $1+1$-dimensional perfect
operator for $-\partial_t^2 + (-\partial_x^2 + m)^2$ developed in the
previous sections. Then we look at the application of the perfect linear
operator to the nonlinear Model A dynamics.

\subsection{Truncated Perfect Operator}

The perfect operator needs to be truncated to finite range to be used in
numerical simulations. Although the introduction of stochastic CG reduced the
interaction range of the perfect action operator, the operator coefficients
do not terminate in a finite range. Furthermore, they decay slowly along
the $x$ direction, where the coefficient of the 10th neighbor\cite{foot_2}
still has an amplitude of around $1\times10^{-5}$. This makes truncation of
the perfect operator more problematic than in quantum field theories,
where keeping next nearest neighbors is already very good\cite{hn}.

One criterion for truncation is that the magnitude of the discarded
coefficients have to be small. But there are other considerations as
well\cite{bieten1,bieten2}. One would like the operator to satisfy certain
constraints that stipulate the correct behavior of the operator in the
continuum limit. These constraints are in the form of sum rules\cite{hn}.
For the diffusion action above, the constraints in the continuous limit are,
\begin{eqnarray}
\left\{\begin{array}{clcl}
&\sum_{i}\sum_{j} H_{i,j} &=& \mu^2 \\
\frac{1}{2}&\sum_{i}\sum_{j}H_{i,j} \,j^2 &=& -r^2 \\
\frac{1}{2}&\sum_{i}\sum_{j} H_{i,j} \,i^2 &=& -2\mu \\
\frac{1}{4!}&\sum_{i}\sum_{j} H_{i,j}\,i^4 &=& 1\,,
\end{array}\right.
\label{eqn:c5_3constraint}
\end{eqnarray}

\noindent
where, as defined previously, $\mu = m\,\Delta x^2$ and
$r = \Delta x^2/\Delta t$. Naively, one might expect that one way of
proceeding would be to truncate the perfect operator to a finite and
manageable range, and then to enforce these constraints to improve the
directly truncated operator. In reality, these sum rules are not
satisfied even for the perfect operator for finite $\Delta x$ and finite
$\kappa$. The error is of higher order in $\Delta x$ and inversely
proportional to $\kappa$. On the one hand, the continuous limit constraint
conditions can be recovered. On the other hand, for finite $\Delta x$, the
constraints no longer hold unless an operator with a long interaction range
is used. The average constraint error is about $0.1\%$ if one keeps up to
$\sim 20$ and $\sim 3$ neighbors in the $x$ and $t$ directions respectively.

\begin{table}
\caption{Coefficients of naturally truncated $11 \times 3$ perfect action
operator for diffusion equation. $\kappa = 2$,
$\mu \equiv m \Delta x^2=0.25$, $\Delta x^2/\Delta t = 1$.}
\begin{center}
\begin{tabular}{|c|c|c|c|c|c|c|c|}\hline
$(t, x)$ & $H$ & $(t, x)$ & $H$ & $(t, x)$ & $H$ & $(t, x)$ & $H$ \\ \hline
(0, 0) &4.00869 &(0, 1) &-1.00198 &(0, 2) &-0.0819891 &(0, 3)
&0.0724608\\ \hline
(0, 4) &0.0270167 &(0, 5) &$3.546940 \times 10^{-4}$ &(0, 6) &-0.002564
&(0, 7) &$-7.593954 \times 10^{-4}$\\ \hline
(0, 8) &$4.298800 \times 10^{-5}$ &(0, 9) &$8.703907 \times 10^{-5}$
&(0, 10) &$1.817881 \times 10^{-5}$ &(1, 0) &-0.430984\\ \hline
(1, 1) &-0.265854 &(1, 2) &-0.046095 &(1, 3) &0.021651 &(1, 4)
&0.012848\\ \hline
(1, 5) &0.001211 &(1, 6) &-0.001157 &(1, 7) &$-4.724628 \times 10^{-4}$
&(1, 8) &$-6.542402 \times 10^{-6}$\\ \hline
(1, 9) &$4.851149 \times 10^{-5}$ &(1, 10) &$1.406475 \times 10^{-5}$
&(2, 0) &$3.220461 \times 10^{-4}$ &(2, 1) &$-2.491026 \times 10^{-4}$
\\ \hline
(2, 2) &$-5.120127 \times 10^{-4}$ &(2, 3) &$1.844663 \times 10^{-4}$
&(2, 4) &$5.223556 \times 10^{-4}$ &(2, 5) &$2.386099 \times 10^{-4}$\\ \hline
(2, 6) &$-2.442266 \times 10^{-5}$ &(2, 7) &$-5.894465 \times 10^{-5}$
&(2, 8) &$-1.676697 \times 10^{-5}$ &(2, 9) &$3.755505 \times 10^{-6}$
\\ \hline
(2, 10) &$4.507152 \times 10^{-6}$ & & & & & &\\ \hline
\end{tabular}
\end{center}
\label{tbl:c5_3rho}
\end{table}

\begin{figure}[tbh]
\begin{center}
\leavevmode
\hbox{\epsfxsize=0.8\columnwidth \epsfbox{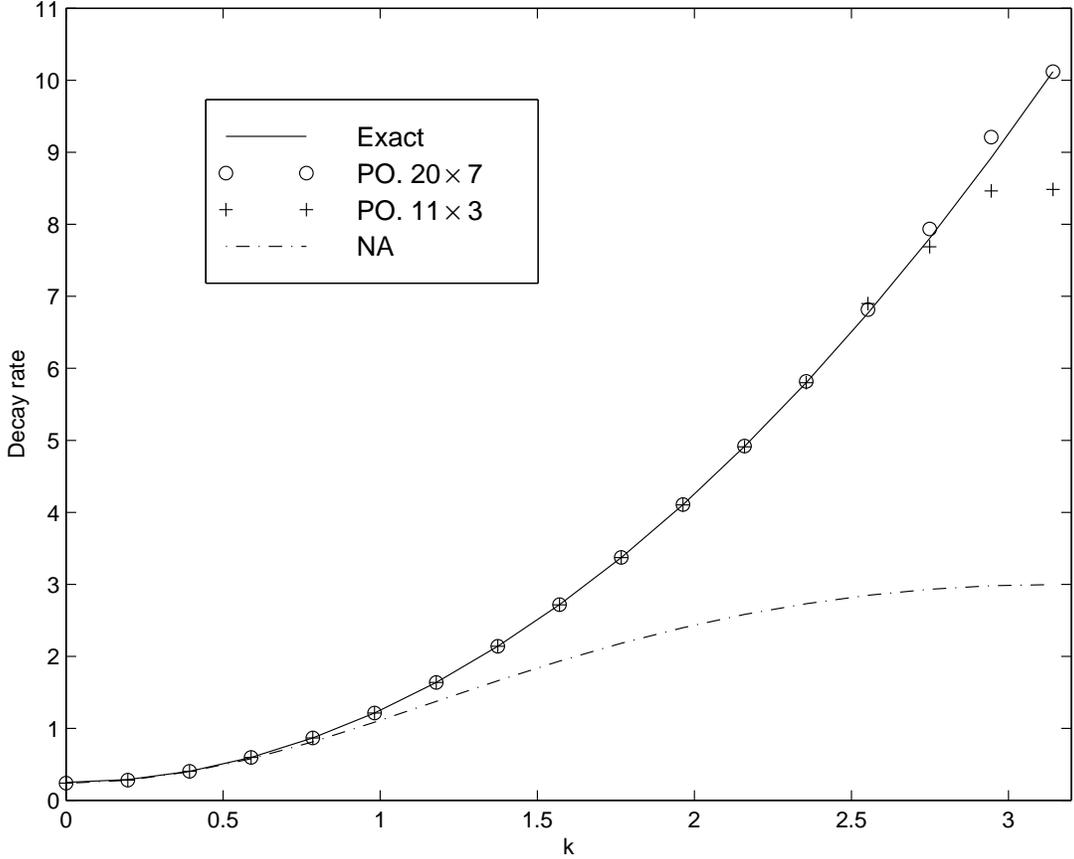}}
\end{center}
\caption{Decay rate of wave modes for diffusion equation. $\kappa=2$,
$\mu=0.25$ and $\Delta x^2/\Delta t=1$. Perfect operator decay rates are
obtained using the first two $t\neq 0$ nodes
(equation (\ref{eqn:c4_4decay})). }
\label{fig:c5_3dispersion}
\end{figure}

An alternative approach\cite{bieten1,bieten2} is to compute the perfect
operator on a smaller lattice and then use this `naturally' truncated perfect
operator. In this way, the constraint is taken care of in the continuum
limit. If we use the operator on a lattice of the same size, the operator
gives a perfect dispersion relation. However, when it is used on a larger
lattice, it is no longer perfect, as can be seen from the inexact dispersion
relation for high wave number modes, which are those most affected by
truncation. The reason lies in the high decay rate associated with a $k^2$
dispersion relation. For $k = \pi$, the ratio between successive $S(k, t)$
values is about $2\times 10^4$. Thus, to maintain exponentially decaying
scaling over three nodes, we need a relative accuracy of $10^{-8}$. Taking
into consideration the importance of keeping enough neighbors and the
computational efficiency, an operator with up to 10th and 2nd neighbors in
the $x$ and $t$ directions is chosen as the operator for most of the
subsequent computer simulations. An operator with 9th and 2nd
neighbors in the $x$ and $t$ directions is also used for some of the
simulations. There is no discernible difference between this operator and
the $11 \times 3$ one.

The operator coefficients are displayed in Table \ref{tbl:c5_3rho} for
$\mu=0.25$ and $\Delta t = \Delta x^2$. For the above operator with
$\mu=0.25$, a 3 node scaling regime is maintained for 60$\%$ of the $k$ mode
and a 2 node scaling regime for about 94$\%$ of $k$ mode. For a larger
operator of size $20\times 7$, we would have a 3 node scaling regime for
about 90$\%$ of the $k$ mode. The decay rates for different operators are
compared in Figure \ref{fig:c5_3dispersion}.

The rapid decay rate of high wavenumber modes is what distinguishes the
perfect operator for the diffusion equation from the $1+1$-dimensional
Laplacian operator used in high energy physics. In the latter case, the
ratio between successive $S(k, t)$ values is at the more benign level of
about $0.04$. The exponential decaying range spans more values of time
displacement. It is easier, therefore, to read off the dispersion relation
all the way to the edge of the Brillouin zone. It is also more stable with
respect to small changes in coefficients of the operator.

\subsection{Numerical Simulation of Diffusion Equation}

We carried out space-time Monte Carlo simulations to test the efficacy of
the perfect operator developed in the previous section. Suppose we are
interested in the diffusion dynamics of the system described by equation
(\ref{eqn:c5_2difEqn}) and would like to calculate its space-time correlation
function. Let the system be of length $L=16$, with the spatial scale of
interest $l=1$. In the path-integral formulation, the time span of the
system is $T=8$. Both space and time directions have periodic boundary
conditions. The Metropolis algorithm is employed\cite{monte}.

Three simulation runs are presented. One simulation uses a perfect operator
with range of interaction up to 10th and 2nd neighbor in the $x$ and $t$
directions respectively. A lattice of $N_x \times N_t = 32 \times 32$ was
used, corresponding to $\Delta x = 0.5$. The other two simulations were
carried out on 32 $\times$ 32 and 64 $\times$ 128 lattices using the
conventional central difference operator. In each case, the time direction
grid size is $\Delta t = \Delta x^2$. In each simulation, $Nrun$ number of
independent runs were conducted to obtain statistics of measurements, each
run with $N = 5 \times 10^5$ MC steps (one sweep of the system) and one
measurement per 8 steps. $Nrun=6$ and 7 for the 32 $\times$ 32 and 64
$\times$ 128 lattices respectively. The $(k \sim 0, \omega \sim 0)$ modes
have the largest standard error, which is crucially dependent on the lattice
size. The typical percentage standard error of $S(k, \omega)$ for 32
$\times$ 32 lattice is about $1\%$ and $2.5\%$ for PO and NA operators,
while that of 64 $\times$ 128 lattice is $6\%$.

\begin{figure}[tbh]
\begin{center}
\leavevmode
\hbox{\epsfxsize=0.44\columnwidth \epsfysize=0.3\textheight
\epsfbox{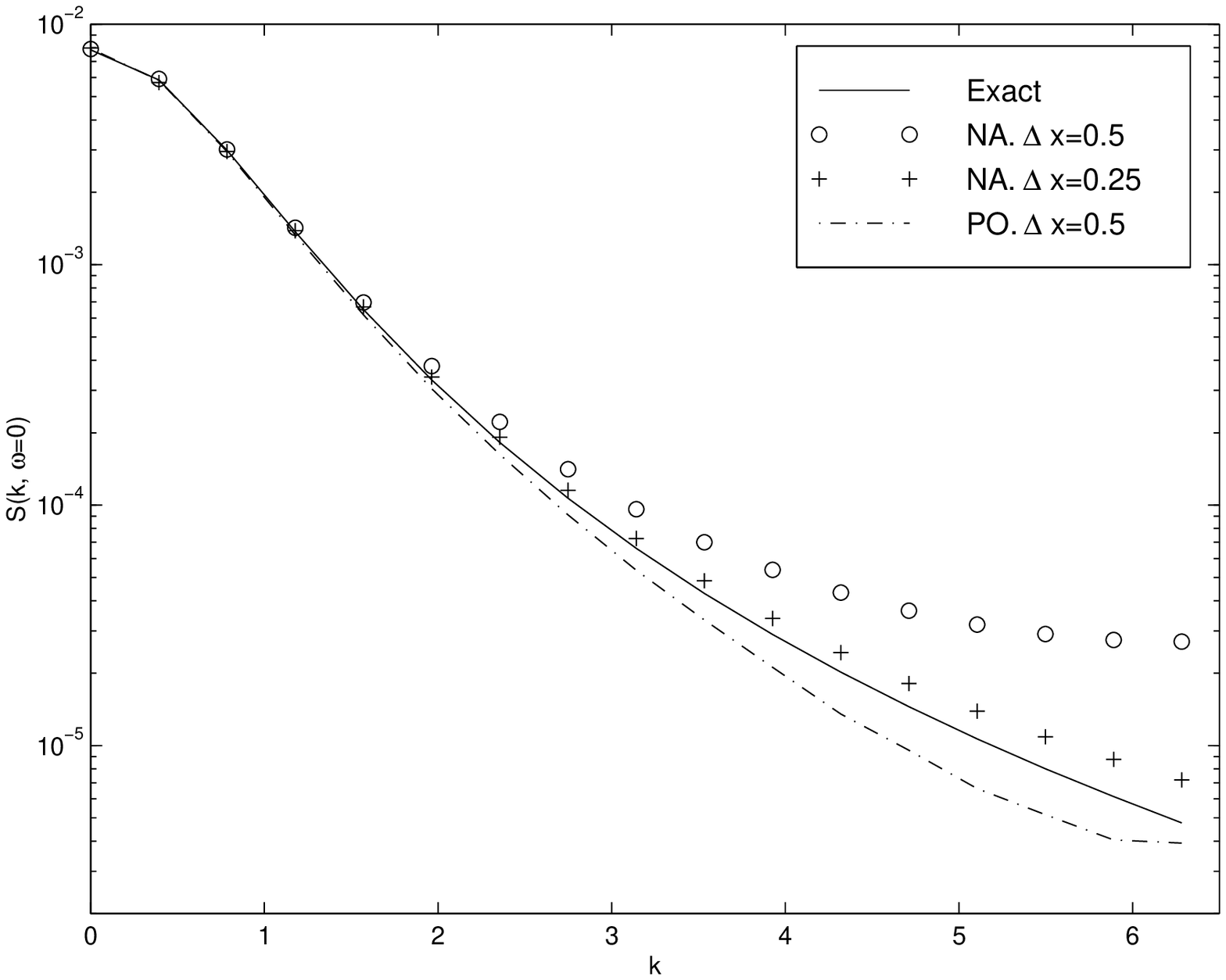}}
\hfill
\hbox{\epsfxsize=0.44\columnwidth  \epsfysize=0.3\textheight
\epsfbox{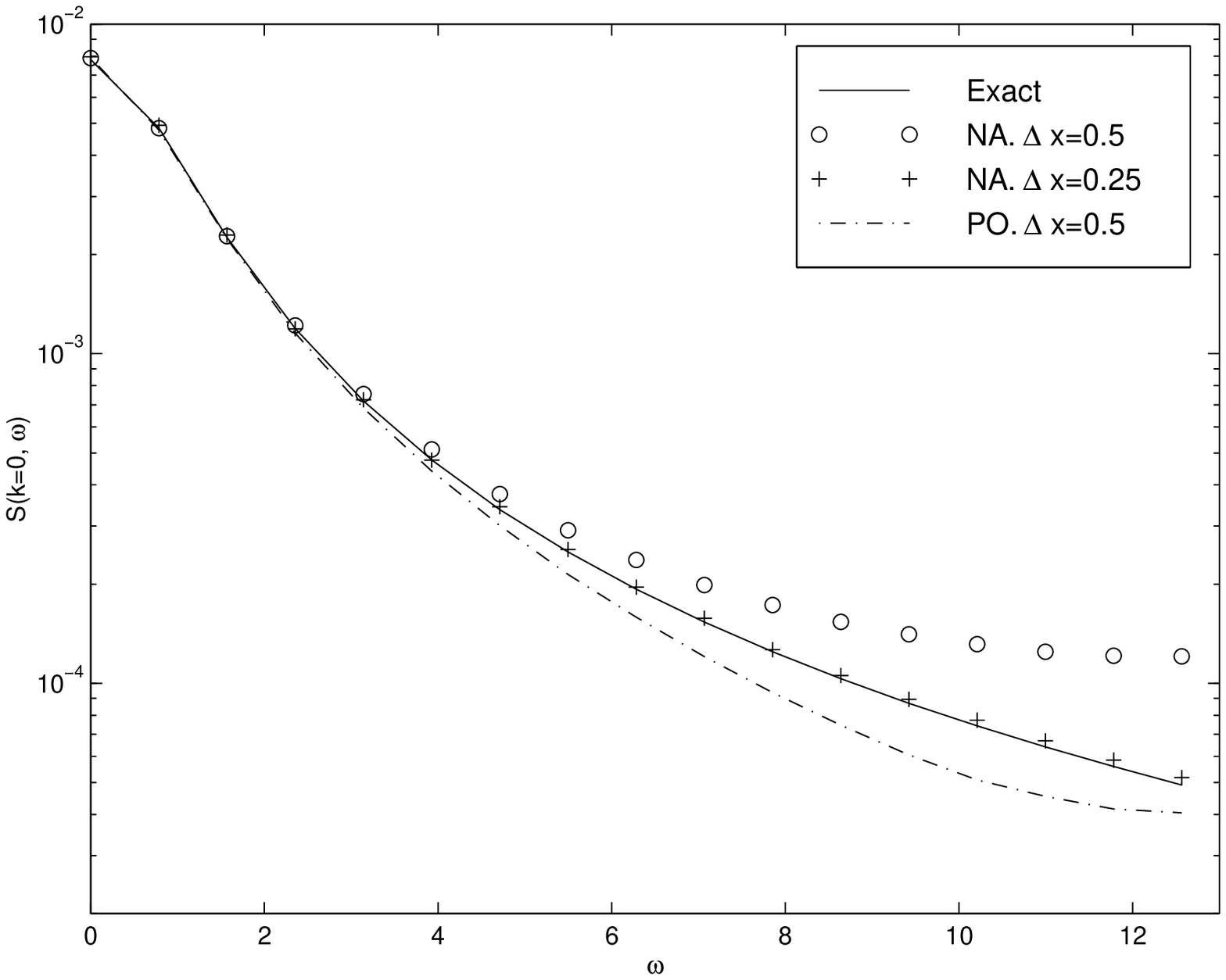}}
\end{center}
\caption{Cross sections of $S(k, \omega)$ for the diffusion equation.
$m=1$, $L=16$, $T=8$. Cross sections are at $\omega=0$ (left) and
$k=0$ (right). The exact result is $[(m + k^2)^2 + \omega^2]^{-1}$.}
\label{fig:c5_4Skw}
\end{figure}

In Fourier space, cross sections of the space-time displaced two point
function $S(k, \omega)$ are plotted in Figure \ref{fig:c5_4Skw}. We do not
expect the perfect operator result to be exact because $S(k, \omega)$ is now
a two point function of the CGed variable, not the continuous variable. But
it turns out to be quite close to the exact result. The NA result for 32
$\times$ 32 lattice deviates further from the true value at the same
$(k, \omega)$ value. For this plot, a constant offset of
$\frac{\Omega}{T\,L}\frac{\Delta x^4}{3\kappa}$ is subtracted from
$S(k, \omega)$ of the perfect operator runs to eliminate the contribution
from the added noise in the stochastic CG transformation.

\begin{figure}[thb]
\begin{center}
\leavevmode
\hbox{\epsfxsize=0.8\columnwidth \epsfbox{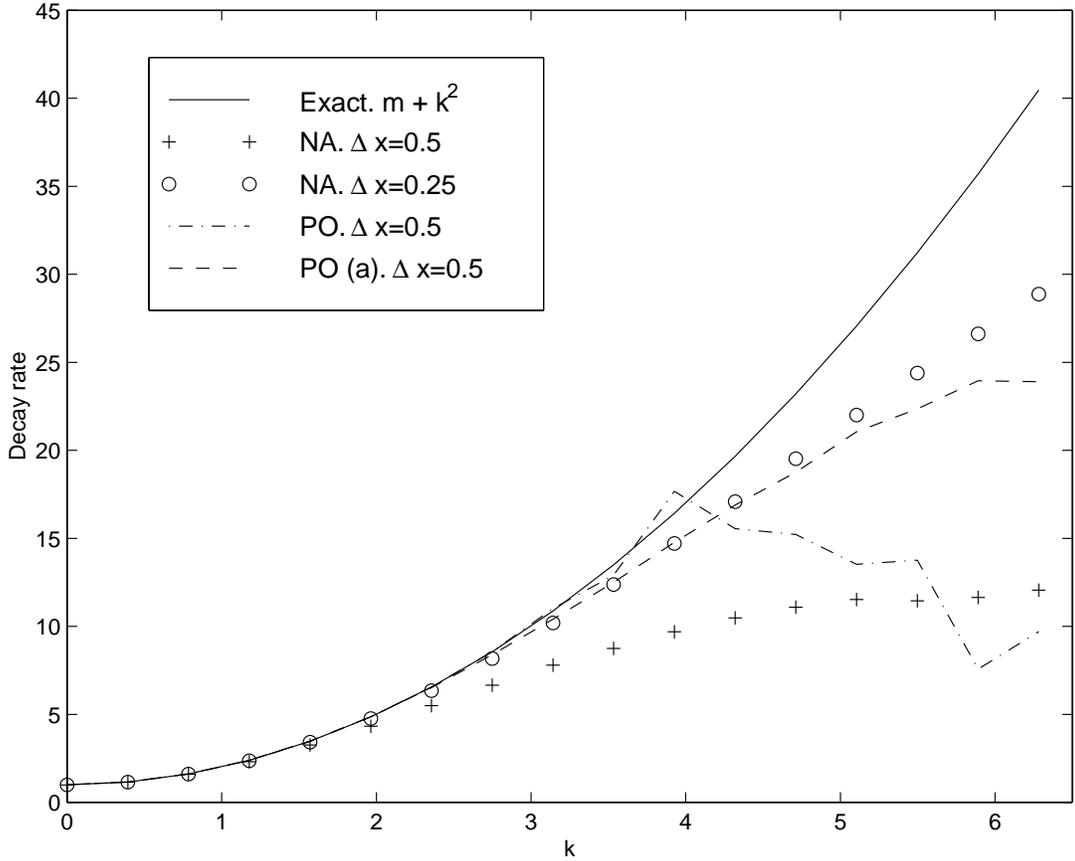}}
\end{center}
\caption{Decay rate of wave modes for the diffusion equation. $m=1$, $L=16$,
$T=8$. Lattices yield $\Delta t = \Delta x^2$. Length scale of interest
corresponds to $k \sim \pi$. Exact result is $m + k^2$. PO results use the
first two $t\neq 0$  nodes of $S(k, t)$. NA results and PO (a) are obtained
using the $t=0$ and $t=\Delta t$ nodes.}
\label{fig:c5_4dispersion}
\end{figure}

Fourier transforming $S(k, \omega)$ to real time, we obtain the dispersion
relation from $S(k, t) \sim e^{-\omega(k) t}$. To avoid static contributions
in the $t=0$ mode, we choose the most significant $t \neq 0$ points to
calculate $\omega(k) = (\log S(k, \Delta t)- \log S(k, 2\Delta t))/\Delta t$.
The results are shown in Figure \ref{fig:c5_4dispersion}. The perfect
operator gives a near `perfect' dispersion relation for the length scale we
are interested in (corresponding to wavenumber $k \sim \pi$), giving the
correct zero $k$ mode mass and correct $k^2$ dependence. We can get a
comparable result using a larger lattice with the NA operator, but with more
computational effort. For large $k$ modes, the amplitude of $S(k, 2\Delta t)$
is of order $10^{-2}$ relative to that at $t=0$ and becomes unreliable given
the simulation accuracy. The real value is used in the plot when
$S(k, 2\Delta t)$ is negative.

One might ask: why call the operator perfect when it does not reproduce the
correct dispersion relation for wavenumbers beyond $k = \pi$? The answer is
that it is not the operator that is not perfect but the simulation itself!
The perfect operator gives the best result possible for physical quantities
of interest given the error of the simulation. With more statistics, the
dispersion relation from the perfect operator approaches the correct result
for all modes with length scale larger than the grid size. The same is not
true for the NA operator. For a discretization twice as fine, with increasing
number of statistical samples, the dispersion relation for the NA operator
approaches a limit that is different from the true solution, and is about
$19\%$ off at the edge of the Brillouin zone.

\begin{figure}[htb]
\begin{center}
\leavevmode
\hbox{\epsfxsize=0.8\columnwidth \epsfbox{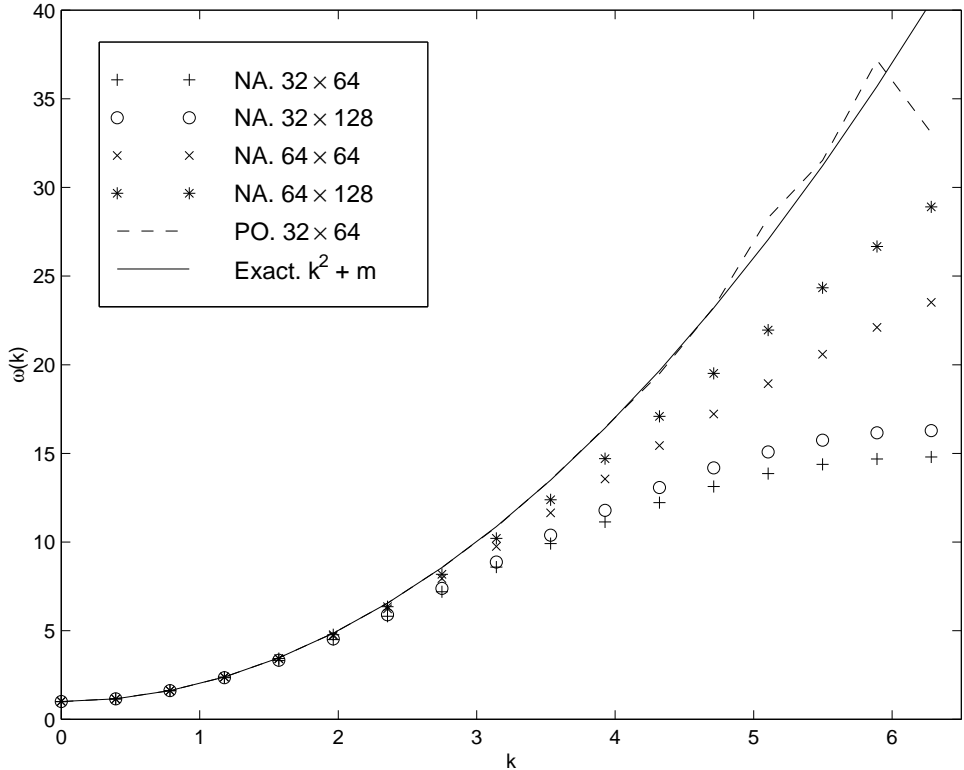}}
\end{center}
\caption{Decay rate of wave modes for the diffusion equation. $m=1$, $L=16$,
$T=8$. Same as in Figure \ref{fig:c5_4dispersion} except that lattices with
$\Delta t \neq \Delta x^2$ are used.}
\label{fig:c5_4dispersion2}
\end{figure}

The simulation error can be overcome when we choose smaller $\Delta t$
relative to $\Delta x^2$. As shown in Figure \ref{fig:c5_4dispersion2}, the
PO decay rates using $\Delta t = \frac{1}{2}\,\Delta x^2$ (corresponding to
$32 \times 64$ lattice) closely follows the exact result and is more accurate
than the measurement from NA.

One might as well choose operators according to the magnitude of the
statistical error of a simulation. Given the usual error of $1\%$ for
$S(k, \omega)$, a smaller sized perfect operator could be used to improve
efficiency of the simulation without compromising accuracy of the physical
measurements. Even with a $11 \times 3$ PO as used in our simulation, the
extra computational effort is not that huge. This operator requires
$21 \times 5 = 105$ points be used to calculate the action density at each
grid point, whereas 7 points are used in conventional NA calculations.
However, since most of the computation effort goes to generating random
numbers (we used Numerical Recipe's ran2() subroutine\cite{NR} as well as
SPRNG modified lagged Fibonacci generator from NCSA\cite{SPRNG}), it turns
out that the overhead from extra neighbors is not significant considering
the improvement of results. If one uses a naturally truncated $5 \times 2$ PO,
total CPU time for the sample calculation will be reduced by $58\%$. The
decay rate rivals the result from NA with a lattice twice as large. In this
case, however, we will not recover a perfect decay rate with better
statistics due to the severe truncation.

Our code is written in C++. On a SUN Ultra2200, the run times are shown in
Table \ref{tbl:c5_4cpuTime} for a test run on a 32 $\times$ 32 lattice with
10000 MC steps. For the same number of steps and lattice size, the PO
calculation takes about 4 times as much time as the NA calculation. Their
standard errors for decay rates are roughly the same if the same nodes are
used. However, the PO uses the second and third nodes to calculate decay
rates. Therefore the resulting decay rates have standard errors about twice
the size of that for NA.

\begin{table}
\caption{CPU time of simulations on diffusion equation using PO vs NA. 32
$\times$ 32 lattice. 10000 Monte Carlo steps. For the same number of
statistical averages, the standard error of $S(k, \omega)$ for PO is about
half of that for NA.}
\begin{center}
\begin{tabular}{|c|c|c|c|}\hline
& Action Calculation & Random Number Generation & Total Time\\ \hline
NA & 6.6s & 15.0s & 30.1s  \\ \hline
PO & 105.4s & 13.6s & 128.5s  \\ \hline
\end{tabular}
\end{center}
\label{tbl:c5_4cpuTime}
\end{table}

The relevant quantity regarding the computational efficiency is the
{\it total computational effort} (TCE) needed to reach a certain level of
root mean square (RMS) error $\delta$. This is defined as
\begin{equation}
\mathrm{TCE} = c\,N\,N_t\,N_x\,,
\end{equation}

\noindent
where the speed factor, $c$, is 4 and 1 for PO and NA respectively. The RMS
error $\delta$ is given by
\begin{equation}
\delta^2 = \delta_1^2 + \delta_2^2\,,
\end{equation}

\noindent
where $\delta_1$ is the bias and $\delta_2$ is the standard error. In
comparing the efficiencies of PO and NA, we focus on the wave mode with
$k = \pi$.

For the naturally truncated PO, $\delta_1 \sim 0.01\%$ and is negligible. For
a $32 \times 64$ lattice, 64K MC steps are needed to reduce $\delta_2$ to
$1\%$ for $k=\pi$. Hence $\mathrm{TCE} = 5.2 \times 10^8$.

For NA and a large lattice size, we have,
\begin{equation}
\delta_1 \approx \frac{a}{N_x^2} + \frac{b}{N_t^2}\,,\ \ {\rm where}\ \
a = \frac{k^4\,L^2}{12\,(m + k^2)}\ \ {\rm and}\ \
b = \frac{(m + k^2)^2\,T^2}{24}\,.
\end{equation}

\noindent
For instance, with $L=16$, $T=8$, $m=1$, $k=\pi$, one has $a = 191.2$ and
$b = 315.1$. The standard error $\delta_2$ is inversely proportional to
$\sqrt{N}$ and is a function of the lattice size. Increasing the lattice
size increases $\delta_2$. However, increasing $N_t$ also has the effect of
improving the result, since smaller $\Delta t$ relaxes the constraint on the
statistical accuracy of the first few nodes of $S(k, t)$. Let us assume
that
\begin{equation}
\delta_2 = \delta_2^{(0)} \,N_x^{\alpha}\,N_t^{\beta}/\sqrt{N}\,,
\label{eqn:c5_4delta2}
\end{equation}

\noindent
where $\alpha$ and $\beta$ are constant parameters. The minimization of the
total computational effort yields,

\begin{eqnarray}
N_x^2 &=& (\frac{2+ 2\,\alpha + 2\,\beta}{2\,\alpha + 1})\,\frac{a}
{\delta_1}\,, \nonumber\\
N_t^2 &=& (\frac{2+ 2\,\alpha + 2\,\beta}{2\,\beta + 1})\,\frac{b}
{\delta_1}\,, \nonumber\\
\frac{N}{N_0} &=& (1 + \frac{1 + \alpha + \beta}{2})\,
(\frac{N_x}{N_{x,0}})^{2\alpha}\,(\frac{N_t}{N_{t,0}})^{2\beta}\,
(\frac{\delta_2(N_0,N_{x,0},N_{t,0})}{\delta})^2\,,
\end{eqnarray}

\noindent
where the optimal $\delta_1 = \delta/\sqrt{1 + \frac{2}{1+\alpha + \beta}}$
and where $\delta_2(N_0,N_{x,0},N_{t,0})$ is the $\delta_2$ value for a
lattice size $(N_{x,0},N_{t,0})$ and with $N_0$ Monte Carlo steps. For
instance, with the above $a$ and $b$ values and $\alpha = \beta = 0$, to
reach a RMS error of $1\%$, one needs $N_x = 257$ and $N_t = 330$. Given that
$\delta_2 \approx 1.2\%$ for $N_0=40K$, $N_{x,0} = 128$ and $N_{t,0} = 256$,
we expect the optimal $N = 86K$. Therefore $\mathrm{TCE} = 7.3 \times 10^9$.
If we have $\alpha = 1$ and $\beta = 0$ instead, the optimal values are
$N_{x} = 190$ and $N_{t} = 423$ and $N = 254K$. Therefore
$\mathrm{TCE} = 2.0 \times 10^{10}$. There is a factor of 40 improvement
(see Table \ref{tbl:c5_4TCE}). The advantage of PO will be more pronounced
in higher dimensions.

The values of $\alpha$ and $\beta$ are difficult to obtain. The values
$\alpha = 1$ and $\beta = 0$ are good approximations for the relevant lattice
sizes, namely $N_x$ and $N_t$ of order of or bigger than 200. Notice that a
large lattice size is most detrimental to the standard error of the small $k$
modes.

\begin{table}[bht]
\caption{Total computational effort for PO vs NA. One requires that the root
mean error of $\omega(k)$ be $\delta = 1\%$ for $k = \pi$. Parameters are
$\alpha = 1$ and $\beta = 0$ (see equation (\ref{eqn:c5_4delta2})).}
\begin{center}
\begin{tabular}{|c|c|c|c|c|c|}\hline
& \ \ \ \ $c$ \ \ \ \ &\ \ \ \ $N_x$ \ \ \ \ & \ \ \ \ $N_t$ & \ \ $N$
($\times 10^3$) \ \ & TCE ($\times 10^8$)\\ \hline
NA & 1 & 190 & 423 & 254 & 200  \\ \hline
PO & 4 & 32 & 64 & 64 & 5.2 \\ \hline
\end{tabular}
\end{center}
\label{tbl:c5_4TCE}
\end{table}

\begin{figure}[htb]
\begin{center}
\leavevmode
\hbox{\epsfxsize=0.8\columnwidth \epsfbox{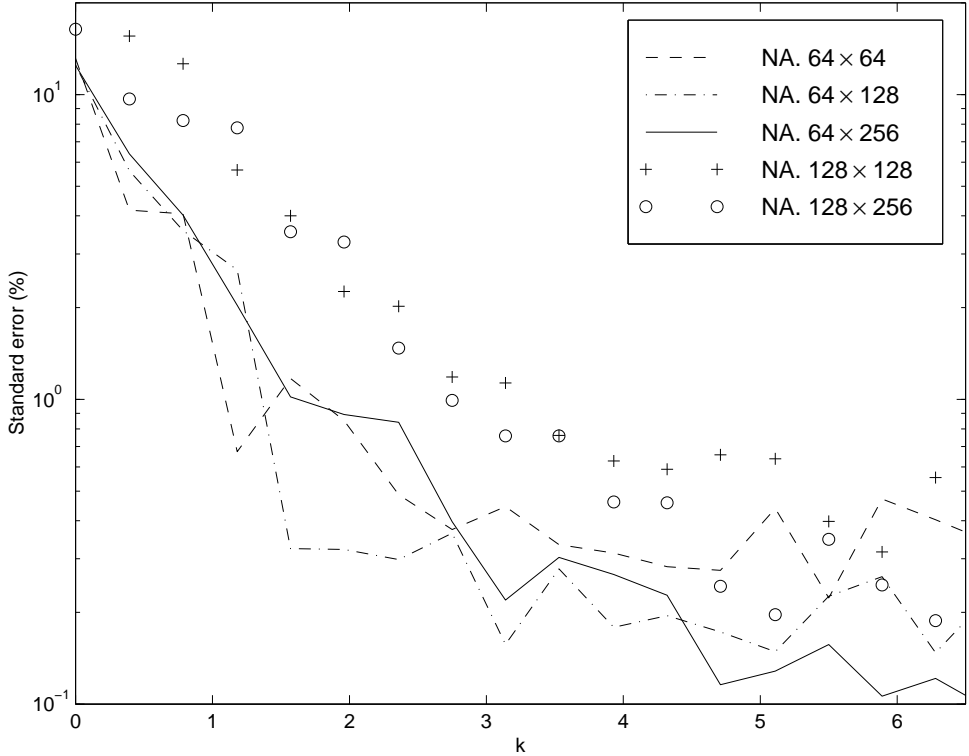}}
\end{center}
\caption{The standard error of the decay rate of wave modes for the diffusion
equation for NA using different lattice sizes. $m=1$, $L=16$, $T=8$. Standard
errors are normalized to $N = 10^5$ Monte Carlo steps.}
\label{fig:c5_4alphabeta}
\end{figure}

In summary, we find that the perfect linear operator gives us the perfect
dynamics of the various wave modes, given the errors of a numerical
simulation. For the same lattice size and number of Monte Carlo steps, the
PO scheme (with the $11 \times 3$ operator) is about 4 times slower relative
to the NA scheme, where generating random numbers takes about $50\%$ of the
total computation time in the latter case. However, the computational effort
in order to reach the same root mean square error for PO is on the order of
$1/40$ of that for NA. This will be more pronounced in higher dimensions.
Moreover, a more severe truncation of the perfect operator is possible, given
the inherent accuracy of the simulation, further enhancing the efficiency of
the PO scheme.

\subsection{Numerical Simulations on Model A Dynamics}

In this section we study the application of perfect linear operator to the
time dependent Ginzburg-Landau equation for Model A dynamics,
\begin{equation}
\partial_t \,\phi = \partial_x^2 \,\phi - m \,\phi - g \,\phi^3  + \eta\,.
\label{eqn:c6_1nonlinearEqn}
\end{equation}

\noindent
The corresponding path-integral formula is,
\begin{equation}
P = \int  D\phi \, \exp\{-\frac{\Delta V}{2 \Omega}\sum_{i, j}
[S_0 + S_1]\}\,,
\end{equation}

\noindent
where $S_0 = \phi(-\partial_t^2 +  (-\partial_x^2 + m)^2)\phi$ and
$S_1 = 2\,g\,\phi^3\,(-\partial_x^2 + m)\phi +
(g\,\phi^3)^2 - \frac{3g\,\Omega}{\Delta x} \phi^2$ are contributions from
the linear and nonlinear terms respectively.

For systems with nonlinear interactions, an exact analytical expression for
the perfect operator is not available. The difficulty lies in the fact that
the form of the continuous action is not closed under the CG transformation.
New interaction terms are generated in reaching the fixed point of the
discrete description of the dynamics. In general there is an infinite number
of interaction terms of diminishing importance. In order to proceed, we need
to make some approximations. In conventional numerical analysis, the form of
the continuous action is used, where the Laplacian operator is replaced by
the central difference operator and local self-interactions are left
unchanged. In analogy, we use the perfect linear operator developed
previously for $S_0$, while leaving the nonlinear self interactions
unchanged. We bundle the $m\phi$ term in with the $g\,\phi^3$ term in
the $m<0$ regime to reduce the standard error of the numerical simulation.
Intuitively this is a reasonable thing to do since $|\phi|$ develops a
non-zero amplitude and the contribution to the dynamics of $\phi$ from these
two terms largely cancel each other. We used the conventional central
difference operator for the operator $-\partial_x^2 + m$ in $S_1$.

There are two regimes: $m>0$ where the nonlinear term amounts to a
renormalization of the mass, and $m<0$ where a nontrivial ground state
develops with a magnitude $\pm \sqrt{m/g}$.

\subsubsection{The $m>0$ Regime}

We simulated the dynamics of a system of physical lengths $L=16$, $T=8$ and
parameters $m=g=\Omega=1$ on lattices of different sizes. Mass dependent
perfect linear operators are used. The Fourier transformed space-time
correlation functions $S(k, \omega)$ are measured and averaged over several
runs. Most simulations consist of $Nrun=9$ runs, each with $N=3\times10^5$
Monte Carlo steps. Measurements are done every 8 Monte Carlo steps. For the
NA result with $64 \times 256$ lattice, 8 runs are used. Fourier transforming
$S(k, \omega)$ to real time, we obtain $S(k, t)$, where $S(k, t=0)$
is the static structure factor and the mode decay rates can be read off from
the time dependence of $S(k, t)$. The length scales of interest are those
larger than $\Delta x = 1$. As in the case of the diffusion equation, the
standard error of the PO result is half of that for NA with the same number
of statistical averages.

\begin{figure}[htb]
\begin{center}
\leavevmode
\hbox{\epsfxsize=0.8\columnwidth \epsfbox{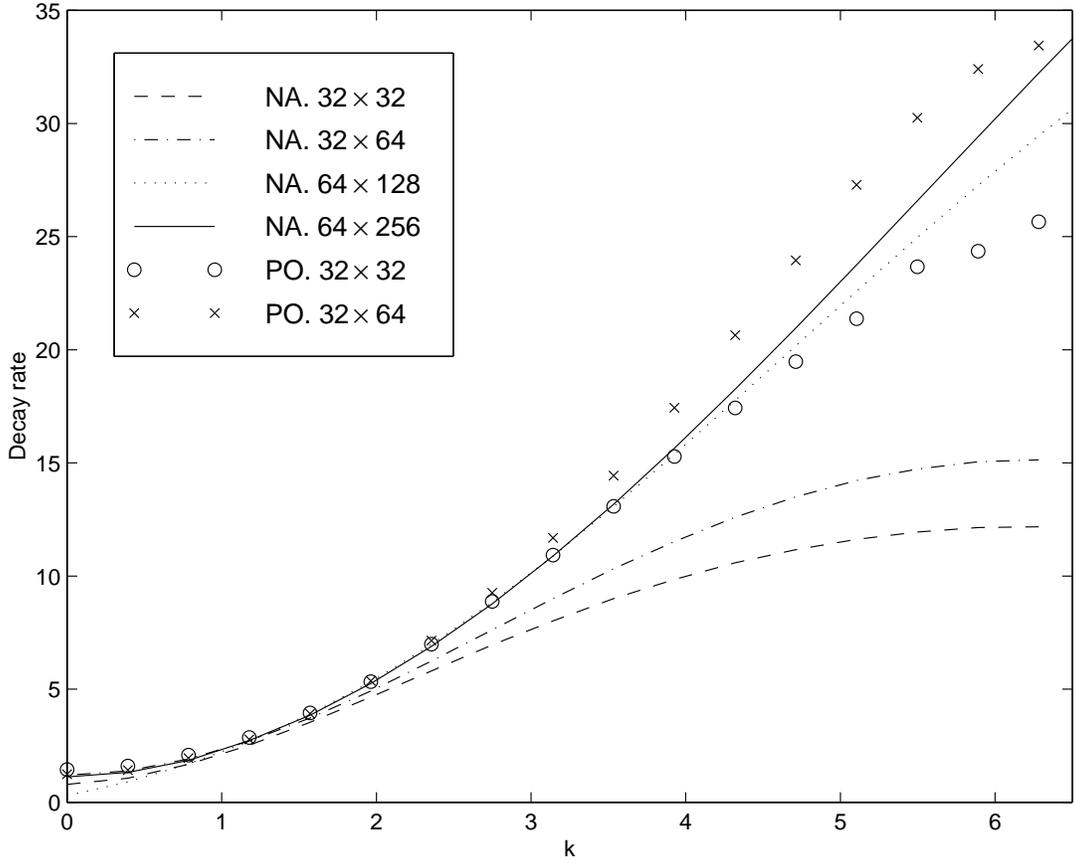}}
\end{center}
\caption{Decay rates of wave modes for the Ginzburg-Landau equation. $m=1$,
$L=16$, $T=8$.}
\label{fig:c6_1AboveX}
\end{figure}

Mode decay rates obtained from the PO scheme for $k$ away from the origin
are greatly improved over its NA counterparts, as shown in
Figure \ref{fig:c6_1AboveX}. For $\Delta x=0.5$, $\Delta t=0.25$, if we had
used the second and third nodes of $S(k,t)$, the decay rates for the second
half of the Brillouin zone would not be reliable, reflecting the inherent
numerical error (roughly $1\%$) of the simulation. This is as in the free
field case discussed at the end of the previous section. For the plots, we
used the  $t=0$ and $t=\Delta t$ nodes instead. It is no longer perfect, but
it is within the numerical error of the simulation and gives improved results
as compared with NA. When we choose $\Delta t=0.125$, the error of the
simulation is no longer a limiting factor and the decay rates over the whole
Brillouin zone are recovered using PO. With a smaller $\Delta t/\Delta x^2$
ratio, the time direction becomes more continuous and the decay rate values
are improved for all schemes as expected.

For $m > 0$, the ground state of the order parameter has an expectation
value of zero. The nonlinear self interaction term in equation
(\ref{eqn:c6_1nonlinearEqn}) has the main effect of renormalizing the mass
to a new effective mass $m_{\mathrm{eff}} = m + g\langle\phi^2\rangle$. In
mean-field theory, the expectation value of $\phi^2$ is expressed as a
function of $m_{\mathrm{eff}}$, which is then self-consistently determined by
the relation
\begin{equation}
m_{\mathrm{eff}}/m = 1 + \frac{1}{m_{\mathrm{eff}}/m}\,
(\frac{g\,\Omega}{4\,m^{3/2}})\,.
\end{equation}

\noindent
The renormalized mass is easily seen to be larger than the bare one.

\begin{figure}[htb]
\begin{center}
\leavevmode
\hbox{\epsfxsize=0.8\columnwidth \epsfbox{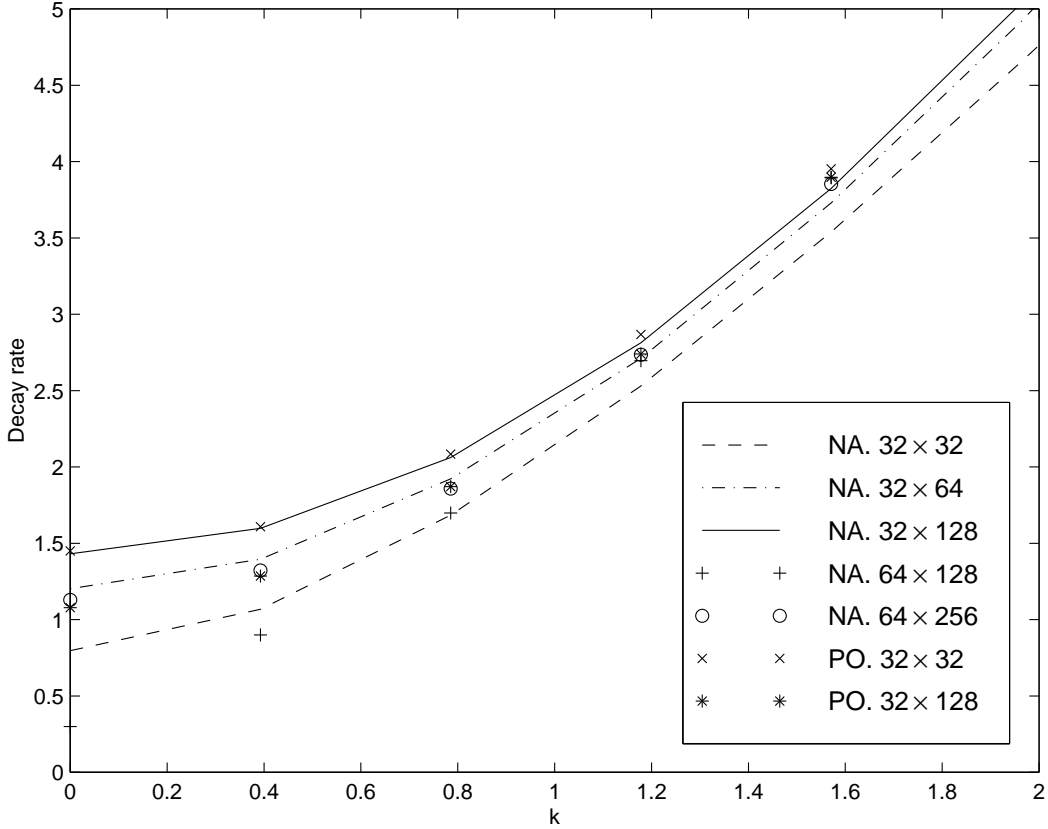}}
\end{center}
\caption{The influence of $\Delta t/\Delta x^2$ on the decay rate of small
$k$ wave modes for the Ginzburg-Landau equation. Smaller $\Delta t$ gives
improved result for NA. The effective mass for PO, however, approaches a
limit less than the mean-field result. $m=1$, $L=16$, $T=8$.}
\label{fig:c6_1AboveXConverge}
\end{figure}

From the decay rate of wave modes with $k\approx 0$, we can read off the
value of the renormalized effective mass. The mean field value of the
effective mass is $m_{\mathrm{eff}}=1.2258$ for the chosen parameters. For
the NA scheme, the renormalized mass is less than the bare mass when the
grid size along the time direction is chosen to be $\Delta t = \Delta x^2$.
Reducing the grid sizes while retaining the ratio $\Delta t/\Delta x^2$ leads
to reduced effective mass values, away from the correct result. For the
$64 \times 128$ lattice, we have $m_{\mathrm{eff}} \approx 0.26$. Unlike in
quantum field theories, time and space are not symmetric in the dynamics we
are considering. This translates into a freedom of choice of grid sizes
$\Delta t$ and $\Delta x$. Physical considerations lead us to the natural
choice of $ \Delta t = c\cdot \Delta x^z$ where $z$ is the dynamic exponent
and $c$ is a constant factor. Outside the critical regime, the diffusion
term dominates the dynamics and $z$ equals the mean field value of 2. We
expect the constant factor $c$ to be dependent on the nature of the
nonlinear interaction and to be different from 1. When we over coarse grain
in the time direction relative to the space direction, the (relatively)
finite size of $\Delta t$ introduces error into the simulation results. We
found that a $\Delta x^2/\Delta t$ ratio value of 2 to 4 is needed to reduce
this error (see Figure \ref{fig:c6_1AboveXConverge}).

\begin{figure}[htb]
\begin{center}
\leavevmode
\hbox{\epsfxsize=0.8\columnwidth \epsfbox{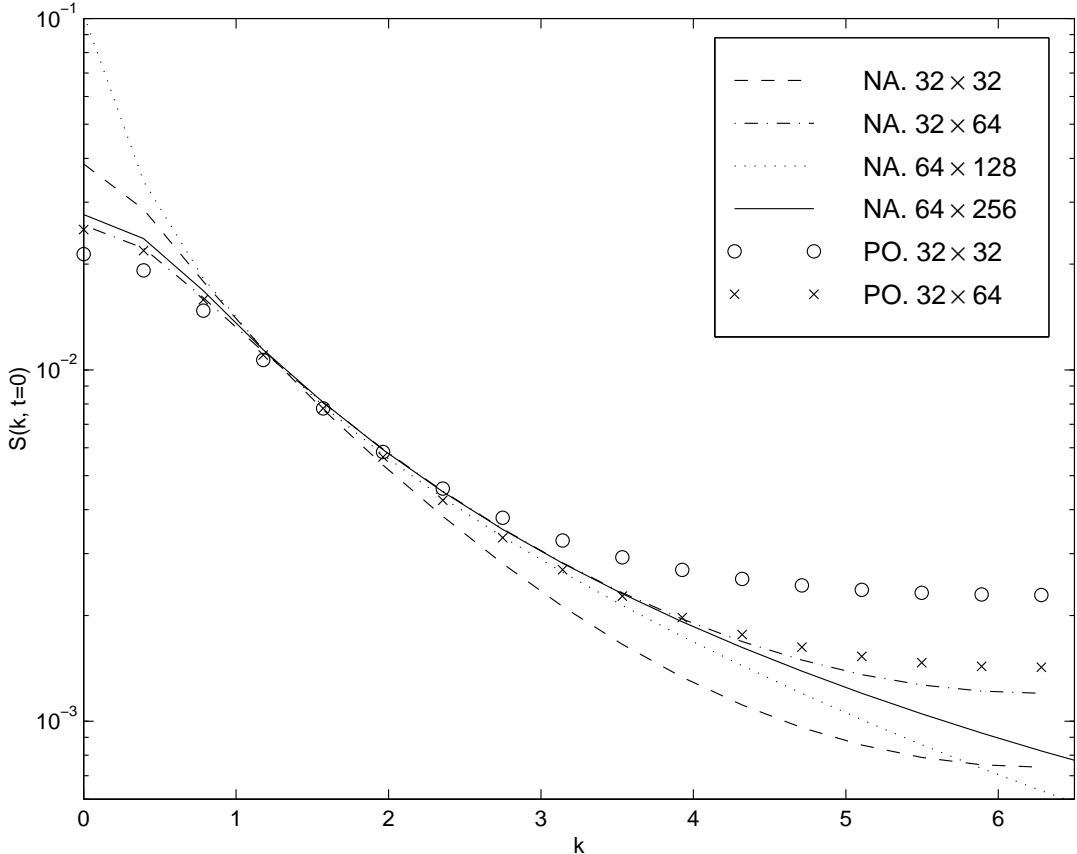}}
\end{center}
\caption{The static structure factor $S(k, t=0)$ for the Ginzburg-Landau
equation. $m=1$, $L=16$, $T=8$.}
\label{fig:c6_1AboveX_t0}
\end{figure}

For the PO scheme, the effective mass is above the bare mass for
$\Delta t = \Delta x^2$. However, as $\Delta t$ is reduced, the effective
mass decreases. For a $32\times128$ lattice, the effective mass is found to be
around 1.07. The reason lies in the fact that we used the simple central
difference Laplacian operator in the nonlinear part of action $S_1$! We
expect that the perfect linear operator operating on a function $f(x)$, which
does not depend on $t$, should yield $(-\partial_x^2 + m)^2\, f(x)$. However,
a summation of the PO along the $t$ direction does not yield the
one-dimensional NA form $(-\partial_x^2 + m)^2$, but rather has coefficients
roughly twice that of the NA form. Therefore, it is inconsistent to simply
use the central difference form for operator $(-\partial_x^2 + m)$. A test
simulation using $\sqrt{2}\,(-\partial_x^2 + m)_{\mathrm{NA}}$ gives the
value 1.36 for the effective mass, closer to our expectation. However, it is
not clear how to interpret this and it points to the need to derive the
perfect form for the whole action including the nonlinear part.

For the static structure factor $S(k, t=0)$, shown in Figure
\ref{fig:c6_1AboveX_t0}, the PO result is not very close to the bench mark
result of NA with a $64 \times 256$ lattice. For large values of $k$, there
is a contribution from the stochastic CG transformation. For small $k$
values, its deviation is a result of the inaccuracy in the effective mass,
which is related to the correlation length $\xi$ (and hence the shape of
$S(k)$) by the relation $\xi \sim m_{\mathrm{eff}}^{-1/2}$.

It is interesting to notice that the structure factor curves obtained using
different schemes and lattice sizes all cross at the same point around
$k \approx 1.3$.

\subsubsection{The $m<0$ Regime}

\begin{figure}[bth]
\begin{center}
\leavevmode
\hbox{\epsfxsize=0.8\columnwidth \epsfbox{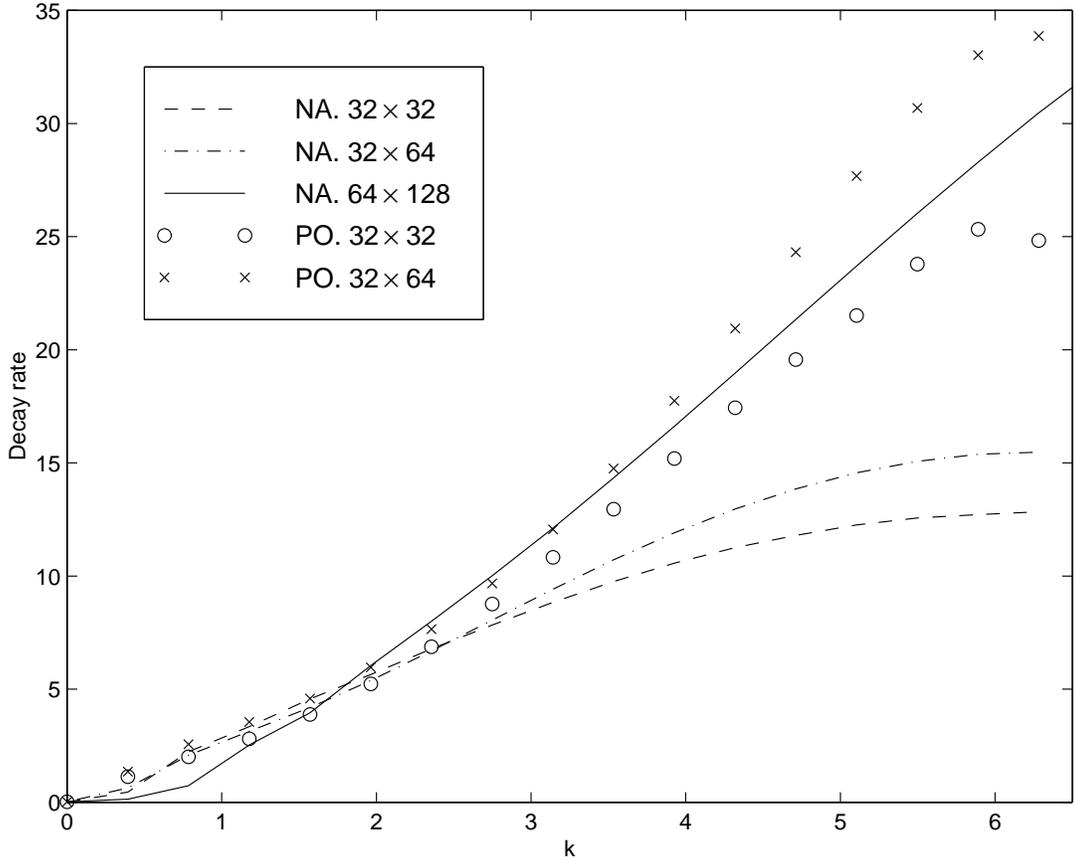}}
\end{center}
\caption{Decay rates of wave modes for the Ginzburg-Landau equation.
$m=-1$, $L=16$, $T=8$.}
\label{fig:c6_1BelowX}
\end{figure}

\begin{figure}[bth]
\begin{center}
\leavevmode
\hbox{\epsfxsize=0.8\columnwidth \epsfbox{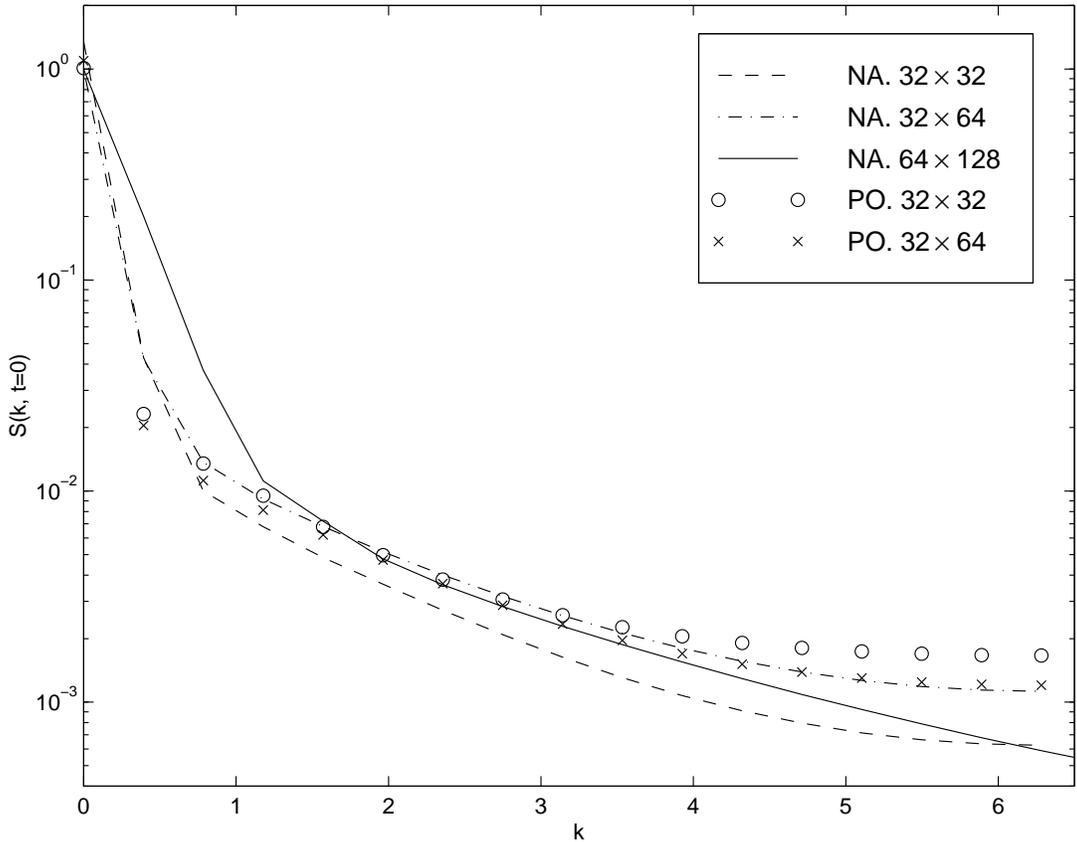}}
\end{center}
\caption{The static structure factor $S(k, t=0)$ for the Ginzburg-Landau
equation. $m=-1$, $L=16$, $T=8$.}
\label{fig:c6_1BelowX_t0}
\end{figure}

In this case, there is a non-trivial fixed point in the action which
corresponds to a ground state with order parameter values
$\phi = \pm \sqrt{m/g}$. Domains of opposite order parameter values compete
and the dynamics is quite different from that with $m$ above 0. In our
simulation, we used the same parameters as in the previous section except
$m=-1$. We treat $m\phi + g\phi^3$ as one term and use the massless perfect
linear operator. This leads to a reduced standard error. The data are
plotted in Figures \ref{fig:c6_1BelowX} and \ref{fig:c6_1BelowX_t0}.

The general shape and values of the dispersion relation are similar to those
of the $m > 0$ regime. However, there is a marked difference between these two
regimes for wave modes close to $k=0$. Here, instead of approaching a finite
effective mass, the decay rate approaches zero, reflecting the existence of
a ground state with a non-zero amplitude. Also due to the `vanishing'
effective mass, the shape of the structure factor is more peaked at the
origin than in the $m > 0$ regime. For modes with small $k$ (first few
nodes), $S(k, \omega)$ values have a large standard deviation. For example,
it is about $25\%$ for the $k = 4\pi/L$ mode and about $9\%$ for the
$k = 8\pi/L$ for NA on a $32 \times 64$ lattice.

When grid sizes are reduced, the dispersion relation changes shape for small
$k$ modes. The difference is significant with respect to the standard error.
This has also been checked with increased statistics. This may be due to
the existence of the non-trivial ground state. For $m < 0$, there is another
length scale in the problem, namely, the interface width between domains
with opposite signs of the ground state order parameter value. If the grid
size $\Delta x$ is not small enough, the position, and hence the dynamics,
of the domain interface will not be resolved. This seems to be the reason why
the shape of the dispersion relation for small $k$ values changes as
$\Delta x$ is reduced, and it places an inherent physical constraint on the
level of discretization one can reach. Only when this extra complication is
taken into account can we obtain a perfect operator for this problem.
Nevertheless, as shown in the figure, the perfect linear operator gives
superior results to the NA operator for the same lattice size and
computational effort (as discussed in the previous section).

In summary, a direct application of the perfect linear operator gives us an
improved dispersion relation for Model A dynamics, especially for those
modes with a length scale comparable to the lattice grid size. However, a
more extensive study is needed to fully assess the efficacy of the perfect
operator. This requires improving the perfect operator such that it yields
the correct effective mass in the $m > 0$ regime and accounts for the
formation of domain interfaces in the $m < 0$ regime.

\subsection{Modified `Perfect' Operator}

As previously shown, although the perfect operator coefficients fall off
exponentially as one moves away from the origin, the decay rate is slow
along the $x$ direction. Therefore, an operator with a shorter range of
interaction is desired.

In non-linear $\sigma$ model\cite{hn}, by simply including the
next-nearest-neighbors (NNN), the dispersion relation can be greatly improved.
In that case, the NNN coefficients are obtained using a natural truncation
of the perfect operator. Since the operator coefficients fall off quickly
along both $x$ and $t$ axis, such a severe truncation can still lead to
significant improvement. This is no longer true for the diffusion equation.
However, we might ask, can we improve the NA operator by allowing for
non-zero operator coefficients for more neighbors? The answer is yes.

\begin{table}[b]
\caption{Coefficients of the modified perfect action operator. $\mu=0$ and
$\Delta x^2/\Delta t = 1$.}
\begin{center}
\begin{tabular}{|c|c|c|c|c|c|c|c|}\hline
$(x, t)$ & $H$ & $(x, t)$ & $H$ & $(x, t)$ & $H$ & $(x, t)$ & $H$ \\ \hline
(0, 0) &6.317206 &(0, 1) &-3.050944 &(0, 2) &$8.922786 \times 10^{-1}$
&(0, 3) &$6.200040 \times 10^{-5}$\\ \hline
(1, 0) &$-4.396585 \times 10^{-1}$ &(1, 1) &$-2.637365 \times 10^{-1}$
&(1, 2) &$-3.045599 \times 10^{-2}$ &(1, 3) &$1.402178 \times 10^{-2}$
\\ \hline
\end{tabular}
\end{center}
\label{tbl:c6_3rho}
\end{table}

We begin from the continuum limit constraints of equation
(\ref{eqn:c5_3constraint}). Setting $\mu=0$ and keeping $\rho(i,j)$ non-zero
for $(i, j)\,\in\, \{(0,0), (0, 1), (1, 0), (2, 0)\}$ (called the basic
points), the conventional operator is obtained as the only solution to these
equations. When more neighbors are included, the constraints are enforced by
solving for $\rho$ of the basic points as function of the other coefficient
values.

\begin{figure}[bh]
\begin{center}
\leavevmode
\hbox{\epsfxsize=0.8\columnwidth \epsfbox{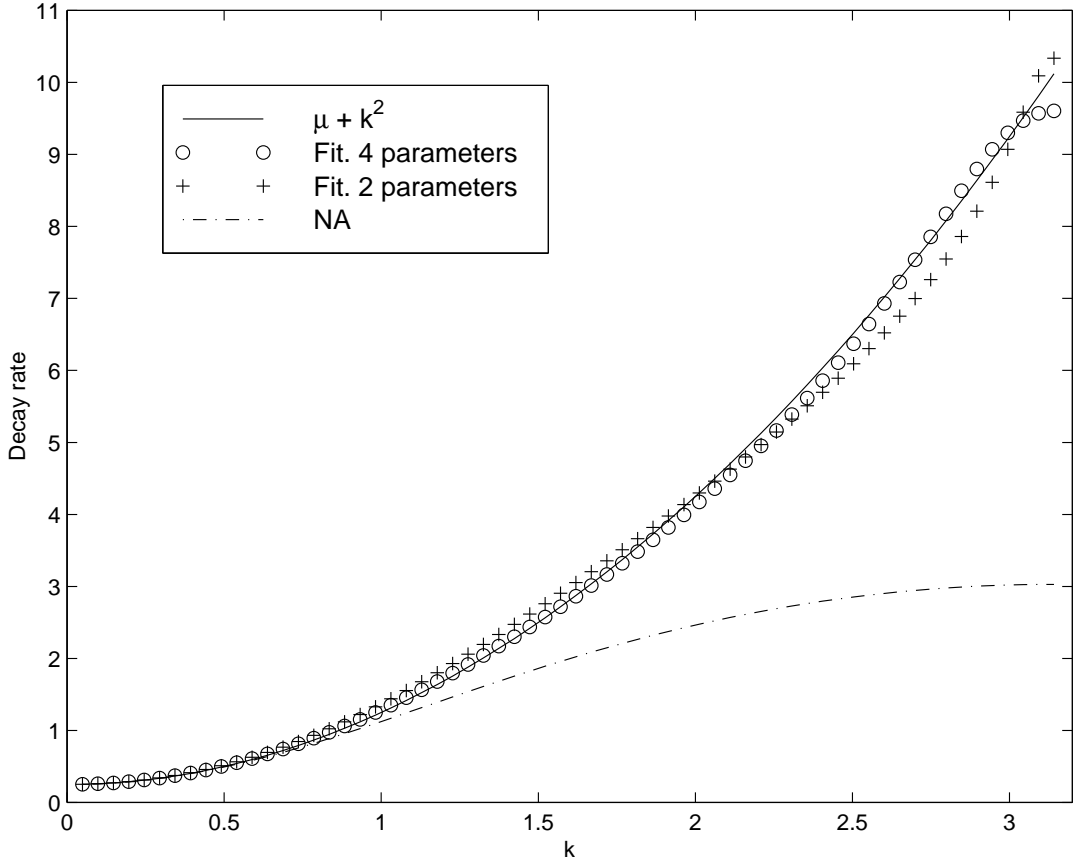}}
\end{center}
\caption{Decay rates of wave modes for the diffusion equation using the
modified perfect operator. $\mu=0.25$, $\Delta t = \Delta x^2$.}
\label{fig:c6_3GeniDispersion}
\end{figure}

Using these non-basic-points coefficients as fitting parameters, we can
obtain an operator with a near perfect dispersion relation. If two parameters
($H(1,1)$ and  $H(2,1)$) are used to obtain a $3 \times 2$ operator, the
average error for the dispersion relation is about $6\%$. By fitting four
parameters (by also including $H(3,0)$ and  $H(3,1)$), we can obtain a
$4 \times 2$ operator --- called the modified perfect operator
(MPO) --- which yields a dispersion relation with an average error of $1.7\%$
with respect to the exact result as shown in
Figure \ref{fig:c6_3GeniDispersion}. The operator coefficients for $m=0$ are
given in Table \ref{tbl:c6_3rho}.

For the MPO, the scaling regime starts from the first time node of the two
point function (i.e. $S(k, t=0)$) due to the nearest neighbor interaction
along the time direction. So long as the first two time nodes have reliable
values, one can estimate the decay rate. This greatly loosens the precision
constraint placed by the perfect operator used before. When the field has
mass, direct fitting under modified constraints that take into account the
mass cause little change in the coefficients.

\begin{figure}[bh]
\begin{center}
\leavevmode
\hbox{\epsfxsize=0.8\columnwidth \epsfbox{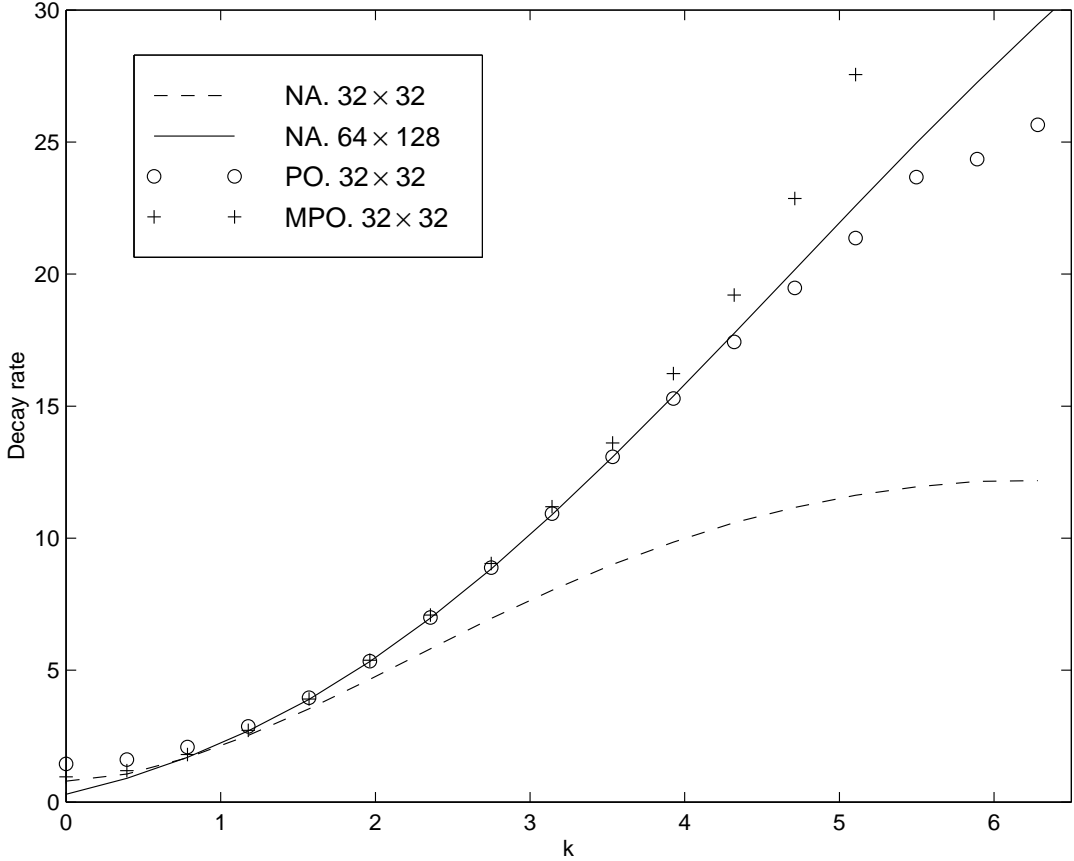}}
\end{center}
\caption{Decay rate of wave modes for the Ginzburg-Landau equation. $m=1$,
$L=16$, $T=8$. The modified perfect operator gives results comparable to
that of the perfect linear operator.}
\label{fig:c6_3AboveXGeni}
\end{figure}

We tested the MPO in simulations of Model A dynamics. The results are
comparable to that of the perfect operator (see Figure
\ref{fig:c6_3AboveXGeni}). It actually gives more accurate decay rates for
wave modes at the edge of the first Brillouin zone, since it allows the use
of the $t=0$ and $t=\Delta t$ nodes to compute the decay rate, while doing
this for the PO is an approximation. The computational effort for the MPO
is drastically reduced due to the relatively short interaction range.

The perfect linear operator operates on the coarse grained variable. For the
modified perfect operator, the physical meaning of the variable it operates
on is not apparent. As discussed in section I, there is a correspondence
between an operator and a specific coarse graining scheme. For the local
averaging CG scheme, or hard CG, the resulting perfect operator has a long
interaction range. However, the range of interaction is reduced after we
modify the CG scheme to be a soft CG scheme dependent on the parameter
$\kappa$. Therefore, it is reasonable to think that there is a variant of
the standard local averaging coarse graining scheme that gives the fast
decaying operator we have computed above. The further investigation of this
point is of general interest as regards the development of an efficient
numerical algorithm.

\section{Conclusions}

The work presented in this paper is a first step towards reaping the
full benefit of using renormalization group in the study of dynamics of
spatially extended systems.  We have constructed perfect
representations of stochastic PDEs that not only integrate out the
small scale degrees of freedom (in space and time), but also develop
non-local representations of the underlying equations that are free of
lattice artifacts.  We demonstrated this by computing the dispersion
relation for elementary excitations, and comparing the results at large
wavenumbers with theoretical expressions valid in the continuum limit.
We exhibited computations for diffusion equations, and a nonlinear
equation derived from model A dynamics, and explored different ways to
truncate the non-local space-time operators generated by the RG.

In one dimension, the computational complexity was reduced by a factor
of about 40 from conventional simulations, for the simple diffusion
problem.  For the nonlinear model A equation, the results were less
impressive, in terms of computer time, because a systematic approximation
scheme for the perfect action has yet to be developed. Nevertheless,
proceeding heuristically, we were still able to obtain improved results
for the static structure factor and the decay rate of modes.  Lastly,
we proposed a heuristic discretization algorithm that incorporates the
ideas of perfect operators, but also gives operators that are more local
than perfect operators.

Finding the perfect operator when nonlinear interactions are present is
a non-trivial task.  The form of the continuous action is not closed
under the CG transformation and new complicated interaction terms are
generated. This is a general property of the RG\cite{nigelbook}. Usually
progress is only possible if the problem under consideration involves a
small parameter which can be used to keep track of the new interactions
which are generated. More generally, the small parameter allows a systematic
approximation scheme to be developed, in which there is a clear prescription
as to which terms have to be included at a given order. If such a parameter
is not available, it is usual to fall back on to some type of variational
scheme, typically including some kind of self-consistent calculation
which corresponds to summing sets of diagrams. Neither of these
approaches have been attempted in this paper. However, we feel that the
results which we have obtained are sufficiently encouraging that some
type of systematic calculation of the perfect operator in nonlinear
theories would turn the ideas presented in this paper into a powerful
computational tool.

\acknowledgments This work was supported in part by the NSF under
grants NSF-DMR-93-14938 and NSF-DMR-99-70690 (QH and NG) and by EPSRC
under grant K/79307 (AM).

\appendix
\section{}

In this appendix, we prove various relations that are important in deriving
the iterative relations for perfect operators.

\begin{enumerate}

\item Here we list some properties of the projection matrices.

We introduced $2N \times N$ matrices $\bar{\hat{R}}$, $\tilde{\hat{R}}$ and
their {\it left} inverses $\bar{\hat{R}}^{-1}= \frac{1}{2}
\bar{\hat{R}}^{T}$, $\tilde{\hat{R}}^{-1}= \frac{1}{2}\tilde{\hat{R}}^{T}$,
\begin{equation}
\bar{\hat{R}}_{m, n} = \delta_{m, 2n} + \delta_{m, 2n-1}, \ \
\tilde{\hat{R}}_{m, n} = \delta_{m, 2n} + \delta_{m, 2n-1} \ \ \
m\in[1, 2N],\ \ n\in[1, N].
\end{equation}

\noindent
Here superscript $T$ indicates transposition. The projection matrices satisfy
relations,
\begin{equation}
\bar{\hat{R}}^{-1}\tilde{\hat{R}} = \tilde{\hat{R}}^{-1}\bar{\hat{R}} = 0
\ \ \mbox{ and } \ \ \bar{\hat{R}}\bar{\hat{R}}^{-1} +
\tilde{\hat{R}}\tilde{\hat{R}}^{-1} = 1\,.
\label{eqn:c3_2rrrr}
\end{equation}

We define the subscripted versions of $\hat{O}$ by
\begin{equation}
\bar{\hat{R}}^{-1}\hat{O}\bar{\hat{R}} \equiv \hat{O}_A
\tilde{\hat{R}}^{-1}\hat{O}\tilde{\hat{R}} \equiv \hat{O}_B
\bar{\hat{R}}^{-1}\hat{O}\tilde{\hat{R}} \equiv \hat{O}_C
\tilde{\hat{R}}^{-1}\hat{O}\bar{\hat{R}} \equiv \hat{O}_D\,,
\end{equation}

\noindent
where $\hat{O}$ and its subscripted versions are linear operators on the
original grid and on the coarse grained grid respectively.

One can prove the following formulae,
\begin{eqnarray}
  g = \hat{O} f \Rightarrow \left\{\begin{array}{l}
\bar{g} =  (\bar{\hat{R}}^{-1}\hat{O}\bar{\hat{R}}) \, \bar{f} +
(\bar{\hat{R}}^{-1}\hat{O}\tilde{\hat{R}}) \, \tilde{f}
\equiv \hat{O}_A\,\bar{f} + \hat{O}_C \,\tilde{f}\\
\tilde{g} =  (\tilde{\hat{R}}^{-1}\hat{O}\bar{\hat{R}}) \, \bar{f} +
(\tilde{\hat{R}}^{-1}\hat{O}\tilde{\hat{R}}) \,\tilde{f} \equiv
\hat{O}_D\,\bar{f} + \hat{O}_B \,\tilde{f}
\end{array}\right.
\label{eqn:c3_2CGg}
\end{eqnarray}

\begin{equation}
{ f^{T}\hat{O} f} = 2\, ({\bar{f}^{T}\,\hat{O}_A\,\bar{f} +
2\bar{f}^{T}\,\hat{O}_C\,\tilde{f} + \tilde{f}^{T}\,\hat{O}_B\,\tilde{f}})
\ \ \mbox{ especially }\ \ f^2 = 2 \,({ \bar{f}}^2 + { \tilde{f}}^2)\,,
\label{eqn:c3_2CGfof}
\end{equation}

\noindent
where we assumed that the matrix $\hat{O}$ is symmetric (physically, this
means $\hat{O}$ possess inversion symmetry), and therefore
$\hat{O}_D = \hat{O}_C^{T}$. Furthermore, if $\hat{O}$ is translational
invariant with $\hat{O}_{m, n} = { \hat{O}}_{m + i, n+i}$, $\hat{O}_A$ and
$\hat{O}_B$ are symmetric while $\hat{O}_C $ and $\hat{O}_D$ are
antisymmetric. This can be seen by looking at their elements,
\begin{eqnarray}
\left\{\begin{array}{ll}
(\hat{O}_A)_{m, n} &= \hat{O}_{2m, 2n} +
\frac{1}{2}(\hat{O}_{2m, 2n+1}+\hat{O}_{2m, 2n-1})\\
(\hat{O}_B)_{m, n} &= \hat{O}_{2m, 2n} -
\frac{1}{2}(\hat{O}_{2m, 2n+1}+\hat{O}_{2m, 2n-1})\\
(\hat{O}_C)_{m, n} &= \frac{1}{2}(\hat{O}_{2m, 2n+1}-\hat{O}_{2m, 2n-1})\,.
\end{array}\right.
\ \ \ m, n = 1,\cdots, N.
\end{eqnarray}

In Fourier space, $\phi(m) = \sum \phi(k) \exp(i k m), \ \ m\in[1, 2N]$ and
$\bar{\phi}(n) = \sum\bar{ \phi}(\kappa) \exp(i \kappa n),\ \ n\in[1, N]$. We
have ${\bar{\hat{R}}}^{-1} = \frac{1}{2} \mathrm{c.c.}\,(\bar{\hat{R}}^{T})$,
${\bar{\hat{R}}}^{-1} = \frac{1}{2} \mathrm{c.c.}\,(\tilde{\hat{R}}^{T})$,
where $\mathrm{c.c}$ is complex conjugate, and
\begin{eqnarray}
\left\{\begin{array}{l}
{ \bar{\hat{R}}}_{k, \kappa}= \sqrt{2}
e^{i\frac{\kappa}{4}}(\cos\frac{\kappa}{4}\delta_{k, \frac{\kappa}{2}} -
i \sin\frac{\kappa}{4}\delta_{k, \frac{\kappa}{2}\pm \pi})\\
{ \tilde{\hat{R}}}_{k, \kappa} = \sqrt{2}
e^{i\frac{\kappa}{4}}(-i\sin\frac{\kappa}{4}\delta_{k, \frac{\kappa}{2}} +
\cos\frac{\kappa}{4}\delta_{k, \frac{\kappa}{2}\pm \pi})
\end{array}\right.
\ \ \ \kappa, k \in (-\pi, \pi).
\label{eqn:c3_2CGsubmFourier}
\end{eqnarray}

\noindent
The sign in $\frac{\kappa}{2} \pm \pi$ should be chosen so that its value
lies within the interval $(-\pi, \pi)$. Physically, equation
(\ref{eqn:c3_2CGsubmFourier}) represents a two step process: folding of the
Brillouin zone by half, such that two wave modes $k$ and $k \pm \pi$ are
mixed,  followed by a stretching back to $(-\pi, \pi)$. This is the
corresponding process in Fourier space of the real space coarse graining
transformation.

Given the operator ${ \hat{O}} = \sum O(k) |k\rangle\langle k|$, i.e. plane
wave functions form its eigenspace, the coarse grained plane waves
$|\kappa\rangle$ are also eigenvectors of ${\hat{O}}_A$, ${ \hat{O}}_B$ and
${ \hat{O}}_C$,
\begin{eqnarray}
\left\{\begin{array}{ll}
(\hat{O}_A)_{\kappa, \kappa'} &= \delta_{\kappa, \kappa'}[\cos^2\frac{\kappa}
{4} O(\frac{\kappa}{2}) + \sin^2\frac{\kappa}{4} O(\frac{\kappa}{2}\pm\pi)]\\
(\hat{O}_B)_{\kappa, \kappa'} &= \delta_{\kappa, \kappa'}[\sin^2\frac{\kappa}
{4} O(\frac{\kappa}{2}) + \cos^2\frac{\kappa}{4} O(\frac{\kappa}{2}\pm\pi)]\\
(\hat{O}_C)_{\kappa, \kappa'} &=  \delta_{\kappa, \kappa'}(-i\cos
\frac{\kappa}{4}\sin\frac{\kappa}{4})[ O(\frac{\kappa}{2})-
O(\frac{\kappa}{2}\pm\pi)]
\end{array}\right.
\ \ \ \kappa, \kappa' \in (-\pi, \pi).
\end{eqnarray}

\item In order to determine the iterative relations of linear operators
(see 3 below), we first have to prove some properties of the subindexed
matrices.

\begin{itemize}
\item[(i)]The first set of properties are:

\begin{eqnarray}
(\hat{O}^{-1})_A \cdot \hat{O}_C &=& - (\hat{O}^{-1})_C \cdot
\hat{O}_B\nonumber\\
\hat{O}_D \cdot (\hat{O}^{-1})_A &=& - \hat{O}_B
\cdot (\hat{O}^{-1})_D\nonumber\\
(\hat{O}^{-1})_D \cdot \hat{O}_C &=& 1 - (\hat{O}^{-1})_B \cdot \hat{O}_B\,.
\label{eqn:app1_0}
\end{eqnarray}

\noindent
To prove the first relation, we use equation (\ref{eqn:c3_2rrrr}):
\begin{eqnarray}
(\hat{O}^{-1})_A \cdot \hat{O}_C
&=&\bar{\hat{R}}^{-1}\,\hat{O}^{-1}\,\bar{\hat{R}} \cdot
\bar{\hat{R}}^{-1}\,\hat{O}\,\tilde{\hat{R}}
= \bar{\hat{R}}^{-1}\,\hat{O}^{-1}\,(1 - \tilde{\hat{R}}\,
\tilde{\hat{R}}^{-1})\,\hat{O}\,\tilde{\hat{R}}  \nonumber\\
&=& \bar{\hat{R}}^{-1}\,\hat{O}^{-1}\,\hat{O}\,\tilde{\hat{R}} -
\bar{\hat{R}}^{-1}\,\hat{O}^{-1}\,\tilde{\hat{R}}\,\tilde{\hat{R}}^{-1}\,
\hat{O}\,\tilde{\hat{R}}\nonumber\\
&=& \bar{\hat{R}}^{-1}\,\tilde{\hat{R}} - (\hat{O}^{-1})_C\,\hat{O}_B
=  - (\hat{O}^{-1})_C\,\hat{O}_B\,.
\end{eqnarray}

\noindent
We can prove the other two relations in a similar way.

\item[(ii)]Another very useful result is:

\begin{equation}
(\hat{O}^{-1})_A = (\hat{O}_A - \hat{O}_C (\hat{O}_B)^{-1} \hat{O}_D)^{-1}
\label{eqn:app1_abcd}
\end{equation}

\noindent
We prove this using equations (\ref{eqn:app1_0}):
\begin{eqnarray}
(\hat{O}^{-1})_A \cdot (\hat{O}_A - \hat{O}_C (\hat{O}_B)^{-1} \hat{O}_D)
&=&(\hat{O}^{-1})_A \cdot \hat{O}_A + (\hat{O}^{-1})_C\,\hat{O}_D\nonumber\\
&=&\bar{\hat{R}}^{-1}\,\hat{O}^{-1}\,\bar{\hat{R}}\,\bar{\hat{R}}^{-1}\,
\hat{O}\,\bar{\hat{R}} + \bar{\hat{R}}^{-1}\,\hat{O}^{-1}\,\tilde{\hat{R}}\,
\tilde{\hat{R}}^{-1}\,\hat{O}\,\bar{\hat{R}}\nonumber\\
&=& \bar{\hat{R}}^{-1}\,\hat{O}^{-1}\,\hat{O}\,\bar{\hat{R}}\nonumber\\
&=& \bar{\hat{R}}^{-1}\,\bar{\hat{R}} = 1
\end{eqnarray}
\end{itemize}

\noindent

\item The iterative relation for the action operator
(equation (\ref{eqn:TheIter2})) follows from that of $\rho$
(equation (\ref{eqn:c4_1IterRho})) which reads

\begin{equation}
(\hat{\rho^{CG}})^{-1} = \hat{L}_C \hat{M}^{-1}
\hat{\rho}_B^{-1}(\hat{M}^{-1})^T \hat{L}^T_C + \hat{\Gamma}\,(\hat{\rho}_A -
\hat{\rho}_C\hat{\rho}_B^{-1}\hat{\rho}_D)^{-1}\,\hat{\Gamma}^T\,,
\end{equation}

\noindent
where $\hat{\Gamma} = \hat{I}+ \hat{O}_C (\hat{O}_B)^{-1}\hat{\rho}_B^{-1}
\hat{\rho}_D$, and where we use $\hat{O}$ to denote the full dynamic
evolution operator $\hat{L}_{\omega}$. Since $\rho$ is symmetric, $A$ and $B$
subindexed matrices are symmetric while $C$ subindexed matrix is the
transpose of $D$ subindexed matrix. Using equations (\ref{eqn:app1_0}) and
(\ref{eqn:app1_abcd}), the second term in the above equation yields,
\begin{eqnarray}
\hat{\Gamma}\,(\hat{\rho}^{-1})_A\,\hat{\Gamma}^T
&=& (\hat{\rho}^{-1})_A - \hat{O}_C (\hat{O}_B)^{-1}\,(\hat{\rho}^{-1})_D -
(\hat{\rho}^{-1})_C\,(\hat{O}^T_B)^{-1}\hat{O}_C^T \nonumber\\
&& + \hat{O}_C (\hat{O}_B)^{-1}\,\left[(\hat{\rho}^{-1})_B -
(\hat{\rho}_B)^{-1}\right]\,(\hat{O}^T_B)^{-1}\hat{O}_C^T\,.
\end{eqnarray}

\noindent
Therefore (if for the US scheme, ignore the factor of 2),

\begin{eqnarray}
(\hat{\rho}^{M})^{-1} &=&(\hat{\rho}^{-1})_A + \hat{O}_C (\hat{O}^{-1})_B\,
(\hat{\rho}^{-1}_B) \,(\hat{O}^T_B)^{-1}\hat{O}_C^T \nonumber\\
&&- \hat{O}_C (\hat{O}_B)^{-1}\,(\hat{\rho}^{-1})_D - (\hat{\rho}^{-1})_C\,
(\hat{O}^T_B)^{-1}\hat{O}_C^T\,.
\end{eqnarray}

\noindent
Therefore, using the iterative relation $(\hat{O}^{M})^{-1} =
(\hat{O}^{-1})_A$ and equations (\ref{eqn:app1_0}), we have,
\begin{eqnarray}
&&(\hat{O}^{M})^{-1} \,(\hat{\rho}^{M})^{-1}\, (\hat{O}^{M})^T\,^{-1}
\nonumber\\
&=& (\hat{O}^{-1})_A\,(\hat{\rho}^{-1})_A\,(\hat{O}^T\,^{-1})_A  +
(\hat{O}^{-1})_A \,\hat{O}_C (\hat{O}_B)^{-1}\,(\hat{\rho}^{-1})_B \,
(\hat{O}^T_B)^{-1}\hat{O}_C^T\,(\hat{O}^T\,^{-1})_A\nonumber\\
&&- (\hat{O}^{-1})_A\,\hat{O}_C (\hat{O}_B)^{-1}\,(\hat{\rho}^{-1})_D\,
(\hat{O}^T\,^{-1})_A - (\hat{O}^{-1})_A\,(\hat{\rho}^{-1})_C\,
(\hat{O}^T_B)^{-1}\hat{O}_C^T \,(\hat{O}^T\,^{-1})_A \nonumber\\
&=& (\hat{O}^{-1})_A\,(\hat{\rho}^{-1})_A\,(\hat{O}^T\,^{-1})_A +
(\hat{O}^{-1})_C\,(\hat{\rho}^{-1})_B \,(\hat{O}^T\,^{-1})_C\nonumber\\
&&+ (\hat{O}^{-1})_C\,(\hat{\rho}^{-1})_D\,(\hat{O}^T\,^{-1})_A+
(\hat{O}^{-1})_A\,(\hat{\rho}^{-1})_C\,(\hat{O}^T\,^{-1})_C\,.
\label{eqn:app1_1}
\end{eqnarray}

\noindent
On the other hand, we have
\begin{eqnarray}
(\hat{O}^{-1}\,\hat{\rho}^{-1}\,\hat{O}^T\,^{-1})_A
&=&   \bar{\hat{R}}^{-1}\,\hat{O}^{-1}\,(\bar{\hat{R}} \,\bar{\hat{R}}^{-1} +
\tilde{\hat{R}}\,\tilde{\hat{R}}^{-1})\,\hat{\rho}^{-1}\,(\bar{\hat{R}} \,
\bar{\hat{R}}^{-1} + \tilde{\hat{R}}\,\tilde{\hat{R}}^{-1})\,
\hat{O}^T\,^{-1}\,\bar{\hat{R}}\,.
\end{eqnarray}

\noindent
Expanding the above equation and comparing with equation (\ref{eqn:app1_1}),
proves equation (\ref{eqn:TheIter2}).

\end{enumerate}


\begin{references}
\bibitem{foot_1}This diagram suggests that coarse-graining and time evolution
commute. This is not true for cases of chaotic dynamics\cite{gmh}.
\bibitem{foot_2}Here a $m \times n$ operator means it has $m \times n$
independent elements. Mirror symmetry with respect to the axis gives us the
other elements in the operator. Along each axis, there are at most $m-1$
and $n-1$ neighbors.
\end{references}
\end{document}